# Random Genetic Variation Links Personality to Earnings and Wealth

Leandro Carvalho (USC)
Sjoerd van Alten (Erasmus University Rotterdam)
Silvia Barcellos (UW-Madison; NBER)
Stephen Dorn (UW-Madison)
Titus Galama (USC; VU Amsterdam)
Catharina A. Hartman (U of Groningen)
Qiongshi Lu (UW-Madison)
Dilnoza Muslimova (VU Amsterdam)

## Abstract

Personality is associated with economic outcomes, but whether these relationships are causal remains unclear. We exploit within-family variation in genetic propensity toward the Big Five generated by random genetic transmission from parents to offspring. Combining polygenic indexes from the Lifelines cohort study with 19 years of administrative tax records in the Netherlands, we find that greater genetic propensity toward emotional stability increases earnings and household wealth, while genetic propensity toward conscientiousness increases household wealth. The other three traits show no consistent pattern across outcomes and specifications. The effect of genetic propensity toward emotional stability on earnings remains positive as we progressively account for genetic propensity toward educational attainment, cognitive performance, and depression. These findings strengthen the evidence that personality causally affects earnings and wealth.

## Introduction

Personality traits predict how lives unfold, but whether they causally affect them is much less clear. The Big Five, a well-established personality framework, captures individual differences along five broad traits: openness to experience, conscientiousness, extraversion, agreeableness, and emotional stability (Goldberg, 1990). These traits predict outcomes ranging from educational and labor market success to health and longevity. This has made personality important not only within psychology but also in other fields including economics (Borghans et al., 2008) and education (Almlund et al., 2011). Furthermore, personality assessments are increasingly used in real-world settings with potential consequences (Brenčič & McGee, 2026). For example, the U.S. Office of Personnel Management identifies personality assessments as tools for employee selection, while the U.S. military administers a personality test to recruits to evaluate their suitability for military life and likelihood of successful adjustment. For some of these applications, prediction may be sufficient.

For many scientific and policy questions, however, the causal effects of personality are central. Psychologists seek to understand the processes through which personality traits shape consequential life outcomes (Hampson, 2021; Roberts & Yoon, 2022). Economists are interested in the role of noncognitive skills, including personality traits (Heckman et al., 2013), in the production of human capital and economic success (Cunha et al., 2010; Cunha & Heckman, 2007; Heckman et al., 2006; Todd & Zhang, 2020). Educational and developmental psychologists have examined the potential for interventions targeting personality-related and socio-emotional skills to support learning (Jach et al., 2023).

However, establishing the causal effects of personality traits on life outcomes is difficult. Personality traits are correlated with other determinants of those outcomes, such as family background, so even rich longitudinal observational studies cannot readily distinguish the effect of personality from those of correlated factors. Previous studies have sought to address this problem by exploiting differences in measured personality between siblings and twins (Fletcher, 2013; Maczulskij & Viinikainen, 2018). More recently, genetic data have provided a complementary source of variation (Hazak et al., 2026; Schwaba et al., 2026). Evidence based on genetic differences across unrelated individuals, however, does not by itself eliminate the problem of confounding: personality-related genetic variation may be associated with family background and other characteristics that also affect those outcomes (Buser et al., 2024).

Here, we exploit within-family variation in genetic propensity toward the Big Five personality traits generated by the random transmission of genetic variants from parents to offspring. Genetic propensity is measured using polygenic indexes (PGIs) for the Big Five—indexes that aggregate millions of genetic variants weighted by their associations with each personality trait. Our design relies on a simple feature of inheritance: at conception, each parent transmits one of their two copies of each variant, and which copy is transmitted is random. By conditioning on parental PGIs, we isolate this random component of genetic inheritance: the remaining individual differences in offspring PGIs are unrelated to family background or other characteristics determined before conception (Benjamin et al., 2026). Big Five PGIs are constructed for approximately 39,000 participants in the Dutch Lifelines Cohort Study using summary statistics from the largest genome-wide association study to date of the Big Five personality traits (Schwaba et al., 2026) and linked to 19 years of administrative records on earnings and wealth.

Together, these advances allow us to estimate the causal effects of genetic propensity toward each of the Big Five personality traits on earnings and wealth in a large within-family sample. Establishing that these effects operate specifically through personality, however, requires addressing pleiotropy. Because genetic variants can influence multiple traits, the same random inheritance that generates differences in genetic propensity toward the Big Five may also generate differences in genetic propensity toward other characteristics that could themselves affect economic outcomes. We investigate the role of these alternative pathways by controlling for PGIs for educational attainment, cognitive performance, and depression—characteristics closely related to personality and economic outcomes.

We find distinct patterns across traits and outcomes. Emotional stability emerges as the clearest result for earnings: greater genetic propensity toward it increases earnings, and the effect remains positive and statistically significant as we progressively control for these pleiotropic pathways. Both emotional stability and conscientiousness increase household wealth. The conscientiousness effect is stable across specifications, whereas the emotional stability effect is sensitive to adjustment for genetic propensity toward depression. The other three traits show no consistent pattern across outcomes and specifications. Although we cannot rule out all pleiotropic pathways, these findings strengthen the evidence that personality causally affects earnings and wealth.

## Data

We link genetic data from participants in the Lifelines Cohort Study, a population-based cohort in the northern Netherlands, to administrative tax records maintained by Statistics Netherlands (CBS). Personality PGIs are constructed for individuals of European genetic ancestry using the GWAS summary statistics from (Schwaba et al., 2026) and standardized to have mean zero and standard deviation one (Supplementary Note 1 and 2). Because few participants have both biological parents genotyped, we use Mendelian imputation to recover the sum of the parental genotypes, leveraging cases in which either one biological parent or one or more full siblings were genotyped (Supplementary Note 3; Young et al., 2022).

Our primary outcomes are annual individual earnings and household net wealth from administrative records, observed from 2006 through 2024. We measure both outcomes in levels and as percentile ranks within birth-year × calendar-year cells, with ranks defined relative to the full Dutch population (Supplementary Note 4). The rank measure is particularly informative for wealth, which is highly skewed and varies substantially over the lifecycle. We restrict observations to ages 30–65 to focus on individuals of working age, after substantial labor market experience has accumulated and prior to retirement.

The resulting sample includes approximately 39,000 individuals. Lifelines used an abbreviated version of the self-report NEO Personality Inventory–Revised (NEO-PI-R) to assess three of the five Big Five traits of its participants: conscientiousness, extraversion, and emotional stability. We use these phenotypic measures in Figure 2 to show that within-family random variation in the corresponding PGIs generates random individual differences in personality traits.

## Research Design

Figure 1 examines the implication of our design that, conditional on the sum of the parental PGIs, individual differences in offspring PGIs should be unrelated to family background. The solid bars in Panels A and B show that an individual's Big Five PGIs are associated with measures of family background, including parental net wealth and number of siblings. The patterned bars show that conditioning on the sum of the corresponding PGIs of the individual's parents substantially attenuates these associations, and we can no longer reject the null of no association. Panel C shows this pattern across a broader set of family-background measures, plotting the absolute *t*-statistics of each association before and after conditioning on parental PGIs. Two of the 45 conditional associations are statistically significant at the 5% level, approximately the number expected by chance.

In our baseline specification, all five Big Five PGIs enter simultaneously, and we condition on the corresponding parental PGIs. Specifically, we estimate:

$$Y_{i,t} = \boldsymbol{Big5}_i \, \boldsymbol{\beta}_1 + \boldsymbol{Big5}_i^P \, \boldsymbol{\beta}_2 + \beta_3 \, Gender_i + \mu_c + \theta_t + \varepsilon_{i,t} \quad (1)$$

where $Y_{i,t}$ denotes the dependent variable for individual $i$ in year $t$; $\boldsymbol{Big5}_i$ is a vector of the individual's Big 5 PGIs; $\boldsymbol{Big5}_i^P$ is a vector containing the sums of the corresponding PGIs of the individual's parents; $Gender_i$ is an indicator for the individual's gender; $\mu_c$ denotes birth-year fixed effects; $\theta_t$ denotes calendar-year fixed effects; and $\varepsilon_{i,t}$ is an error term. Standard errors are clustered at the individual level. Some of the dependent variables are time-invariant. In those cases, the $t$ subscript and the calendar-year fixed effects $\theta_t$ are omitted, and we report heteroskedasticity-robust standard errors.

## Results

Figure 2 shows that within-family random variation in the Big Five PGIs generates variation in the corresponding personality traits. A one-standard-deviation (SD) increase in the conscientiousness, extraversion, and emotional stability PGIs increases their corresponding survey-based personality measures by 0.11 (SE 0.008), 0.09 (SE 0.008), and 0.14 SD (SE 0.008), respectively (all P < 0.001; n = 37,768). Openness and agreeableness are not measured in Lifelines. Consistent with the PGIs capturing trait-specific variation, each PGI has its largest effect on the personality trait it was constructed to predict. Compared with estimates that do not condition on the parental PGIs (Supplementary Materials Table S7), these estimates exhibit little attenuation, consistent with the within-family results reported by (Schwaba et al., 2026).

Figure 3 examines whether this within-family random variation in the Big Five PGIs also generates differences in economic outcomes. It shows distinct patterns across traits and outcomes. Emotional stability emerges as the clearest result for earnings. A one-SD increase in the emotional stability PGI increases annual individual earnings by €1,420 (SE €203, P < 0.001), or 3.3% of the sample mean of €43,385, with a similarly positive effect when earnings are measured in percentile ranks (1.38 percentile points; SE 0.18, P < 0.001). The estimated effects of the other four Big 5 PGIs on earnings are considerably smaller.

For household wealth, both emotional stability and conscientiousness have positive effects. A one-SD increase in the emotional stability PGI increases household net wealth by €6,322 (SE €2,219, P = 0.004), or 3.2% of the sample mean of €195,543, while a one-SD increase in the conscientiousness PGI increases it by €4,344 (SE €2,190, P = 0.047). The percentile-rank estimates tell a similar story, with both PGIs increasing individuals' relative position in the wealth distribution (emotional stability: 0.72 percentile points, SE 0.20, P < 0.001; conscientiousness: 0.53 percentile points, SE 0.20, P = 0.007). The conscientiousness effect is estimated more precisely using wealth ranks than wealth levels. As discussed above, ranks reduce the influence of the highly skewed wealth distribution and compare individuals at similar stages of the lifecycle. Openness, extraversion, and agreeableness show no consistent pattern across wealth.

A challenge to interpreting the effects of the Big Five PGIs as operating through personality is pleiotropy. The solid bars in Panel A of Figure 4 illustrate this possibility: random variation in emotional stability PGIs also generates differences in years of schooling. We investigate this possibility by augmenting equation (1) with controls for the individual's and their parents' educational attainment PGIs (EA PGIs; Okbay et al., 2022), rather than the individual's realized education, which may itself lie on the causal pathway from personality to economic outcomes. Figure 4 compares estimates with and without these controls. Panel A shows that controlling for the EA PGIs substantially attenuates the effects of the conscientiousness and emotional stability PGIs on years of schooling, while Panel B shows that their effects on their corresponding personality traits are nearly unchanged. Thus, controlling for the EA PGIs removes genetic variation shared between personality and educational attainment while preserving genetically induced variation in personality. Panel C shows that conditioning on the EA PGIs attenuates the effect of the emotional stability PGI on earnings by about a third, from €1,420 to €957 (SE €204, $P < 0.001$), but a sizable positive effect remains. The percentile-rank estimate falls similarly, from 1.38 to 0.98 percentile points (SE 0.18, $P < 0.001$). For wealth in Panel D, the effect of the conscientiousness PGI is nearly unchanged, from 0.53 to 0.57 percentile points (SE 0.19, $P = 0.004$), while the effect of the emotional stability PGI is attenuated by about a third, from 0.72 to 0.49 percentile points (SE 0.20, $P = 0.013$).

Supplementary Materials Table S6 extends this analysis by additionally controlling for PGIs for cognitive performance (Lee et al., 2018) and major depressive disorder (MDD) (McIntosh et al., 2024). Supplementary Materials Table S8 reports the relationships among these PGIs and the Big Five PGIs, while Supplementary Materials Table S9 reports the effects of the EA, cognitive-performance, and MDD PGIs on their corresponding measured phenotypes. The effect of the emotional stability PGI on earnings is attenuated but remains positive across specifications (0.66 percentile points with all three sets of controls, SE 0.20, $P < 0.001$). The conscientiousness effect on wealth is stable across specifications (with all three sets of controls: 0.44 percentile points, SE 0.19, $P = 0.025$), while the emotional stability effect on wealth remains positive after controlling for the EA and cognitive performance PGIs (0.53 percentile points, SE 0.20, $P = 0.008$) but is sharply attenuated when the MDD PGIs are added (0.02 percentile points, SE 0.21, $P = 0.92$). The latter finding should be interpreted cautiously because depression itself may be downstream of emotional stability (Jeronimus et al., 2016).

Finally, in Figure 1, we conduct a descriptive, non-causal mediation analysis by examining how the estimated relationships between the conscientiousness and emotional stability PGIs and the economic outcomes change when we add measured personality traits to the regressions one at a time (Supplementary Materials Figure S1-S2 and Table S10). The pattern is trait-specific. Controlling for measured emotional stability reduces the effect of the emotional stability PGI on earnings and wealth, whereas controlling for measured conscientiousness or extraversion does not. Similarly, controlling for measured conscientiousness reduces the effect of the conscientiousness PGI on wealth, whereas controlling for measured emotional stability or extraversion has little effect. Although we do not interpret this exercise as identifying a formal mediation effect, this trait-specific attenuation is consistent with the Big Five PGIs affecting economic outcomes through their respective personality traits.

# Discussion

A large observational literature documents associations between personality and earnings (Alderotti et al., 2023; Vella, 2024) , and a smaller literature documents associations between personality and wealth (Brown & Taylor, 2014; Jiang et al., 2024; Mosca & McCrory, 2016), but distinguishing causal effects from confounding has been a longstanding challenge. Previous studies have sought to address confounding by exploiting differences in measured personality between siblings and twins. (Fletcher, 2013), who examines all five traits, finds that emotional stability and extraversion are positively associated with earnings, while (Maczulskij & Viinikainen, 2018), who measure emotional stability but not the other Big Five domains, find that greater emotional stability is associated with higher earnings.

More recently, studies have used genetic data to examine the relationship between personality and income. Schwaba et al. (2026) find positive genetic correlations of income with emotional stability, agreeableness, extraversion, and openness, while Hazak et al. (2026), who focus on emotional stability, find that its PGI is positively associated with income. Because these estimates rely on genetic variation across unrelated individuals, however, they may still reflect confounding by family background and other correlated characteristics—a concern supported by the associations between personality PGIs and family background documented in Figure 1.

Recent advances make it possible to make further progress on this longstanding question. The large genome-wide association study of Schwaba et al. (2026) provides substantially more predictive PGIs for the Big Five personality traits, while Mendelian imputation (Young et al., 2022) allows us to exploit the randomness of genetic transmission from parents to offspring in a large sample. Unlike designs based on genetic differences between siblings, the approach used is not susceptible to bias from sibling indirect genetic effects (Young et al., 2022). We combine these advances with 19 years of administrative tax records on earnings and wealth.

This combination allows us to establish that random within-family variation in genetic propensity toward emotional stability generates differences in earnings. This finding is consistent with Buser et al. (2024), who find suggestive evidence of relationships between various PGIs for noncognitive skills and labor market outcomes. We also find that greater genetic propensity toward emotional stability and conscientiousness increases household wealth. Our main findings are robust to progressively conditioning on PGIs for educational attainment, cognitive performance, and depression. These findings suggest that the economic consequences of personality extend beyond labor market earnings to the accumulation of household wealth, potentially through channels such as saving and investment behavior. To our knowledge, ours is among the first studies to investigate the causal relationship between personality and wealth.

Our earnings results have practical implications for hiring. Emotional stability receives considerably less attention than the other Big Five traits in interviews and job advertisements (Brenčič & McGee, 2026; Sackett & Walmsley, 2014). Our findings therefore highlight a contrast between the dimensions of personality emphasized in hiring and those that emerge most clearly for earnings in our analysis.

An important limitation is that our approach cannot rule out pleiotropy altogether. The extent to which these controls address pleiotropy depends on how well the PGIs capture genetic variation in potential alternative pathways. For example, conditioning on the highly predictive EA PGI eliminates the relationship between the personality PGIs and schooling while leaving their relationships with measured personality largely unchanged, whereas the less predictive cognitive-performance PGI may capture its corresponding pathway less completely. At the same time, conditioning on PGIs for alternative pathways may remove effects that originate in personality but are mediated through those pathways. If emotional stability affects education, for example, controlling for the EA PGI may remove part of an effect of emotional stability on earnings that is mediated through education. In this sense, our adjusted estimates may understate the total effect of personality on economic outcomes.

Two other limitations concern generalizability and the interpretation of PGIs. Our sample consists of individuals of European genetic ancestry from the Netherlands, and the extent to which these findings generalize to other populations remains uncertain. Moreover, PGIs are context-dependent measures whose weights may vary across populations and environments (Furrer & Turkheimer, 2026) . Throughout, we use the term genetic propensity to describe the random within-family variation in the Big Five PGIs, rather than to imply an immutable or context-independent genetic predisposition. Although we exploit random genetic transmission for identification, this does not imply that the resulting economic effects are biologically predetermined. Genetic differences may ultimately affect earnings and wealth through behavioral and environmental pathways.

Several features of our results strengthen their causal interpretation. We exploit random within-family genetic transmission, our main findings are robust to conditioning on PGIs for educational attainment, cognitive performance, and depression, and the mediation analysis provides suggestive evidence that the effects of genetic propensity operate through the corresponding personality traits. These analyses cannot rule out pleiotropy, but collectively they strengthen the evidence that personality contributes causally to socioeconomic differences rather than merely being correlated with them.

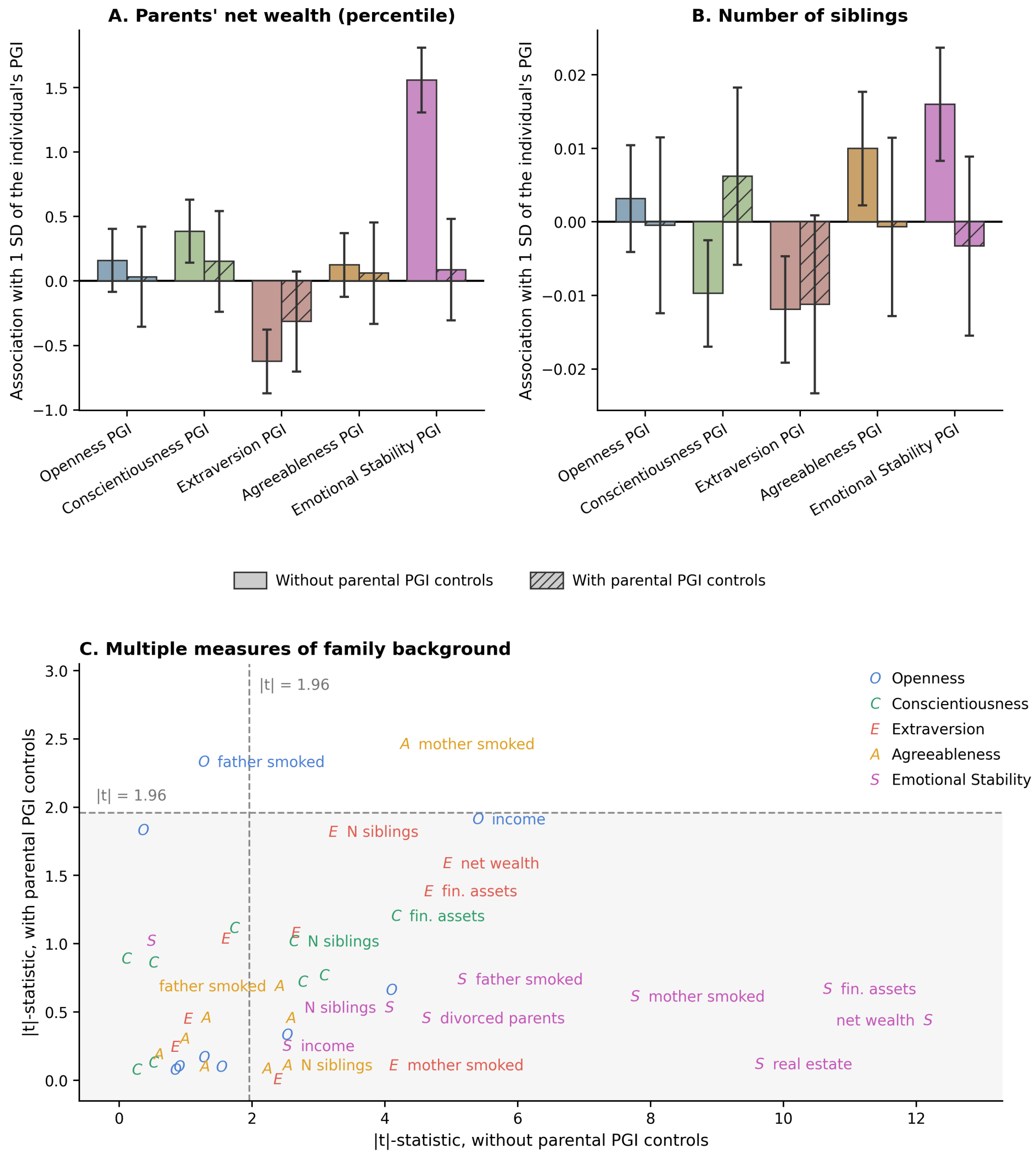

**Fig. 1. Big Five PGIs and Family Background.** The figure examines the implication of our design that, conditional on the sum of the parental PGIs, individual differences in PGIs should be unrelated to family background. Panels A and B report associations between the individual's Big Five PGIs and parents' net wealth and number of siblings, respectively. For each outcome, the five coefficients on the personality PGIs are estimated jointly. Solid bars report estimates without conditioning on parental PGIs, while patterned bars report estimates that condition on the sums of the corresponding PGIs of the individual's parents. Panel C extends this comparison to a broader set of family-background measures, plotting the absolute t-statistics of each association before and after conditioning on parental PGIs. The dashed lines indicate the critical value for statistical significance at the 5% level. Error bars in Panels A and B represent 95% confidence intervals. Number of observations range from 30,367 to 36,823.

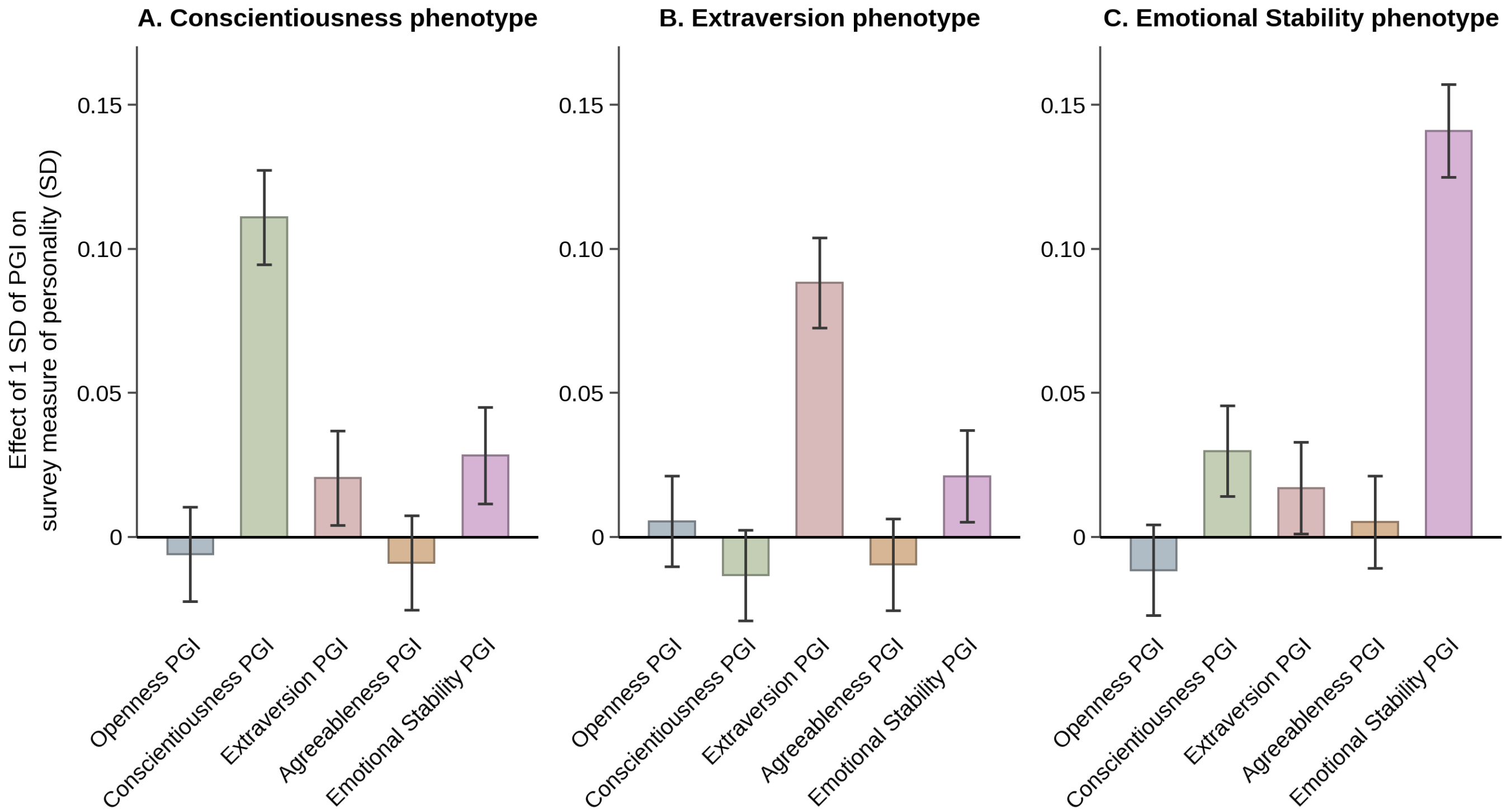

**Fig. 2. Big Five PGIs and Survey Measures of Personality Traits.** The figure examines whether the random variation in the Big Five PGIs generates corresponding individual differences in personality traits. It reports estimates of equation (1), in which the five coefficients on the Big Five PGIs are estimated jointly, conditioning on the sums of the corresponding PGIs of the individual's parents. Panels A, B, and C report estimates for conscientiousness, extraversion, and emotional stability, respectively. PGIs and personality measures are standardized, so coefficients are expressed in standard-deviation units. Error bars represent 95% confidence intervals. The sample includes 37,768 individuals.

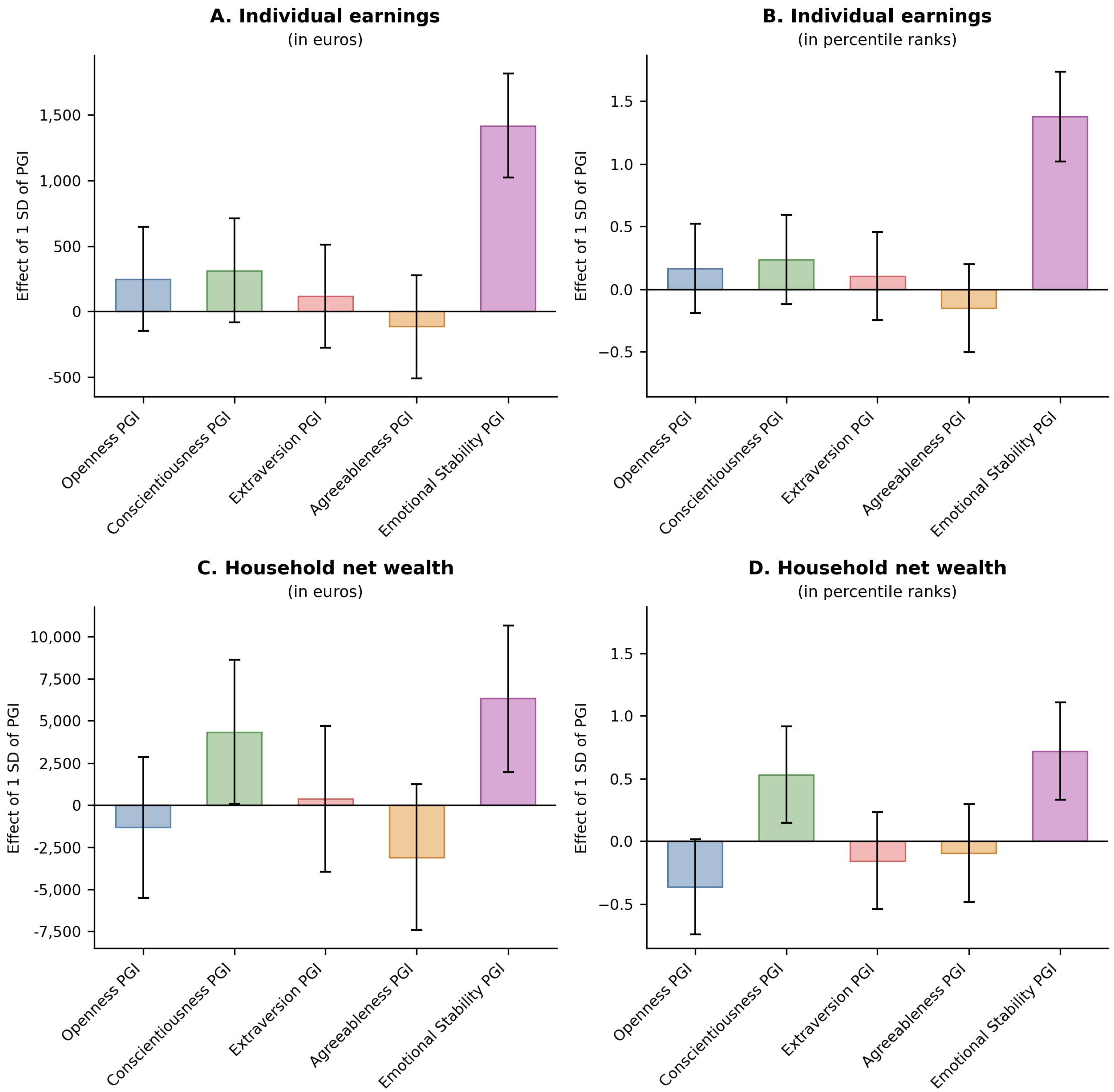

**Fig. 3. Big Five PGIs and Economic Outcomes.** The figure examines whether the random variation in the Big Five PGIs generates corresponding individual differences in economic outcomes. It reports estimates from equation (1), in which the five coefficients on the Big Five PGIs are estimated jointly, conditioning on the sums of the corresponding PGIs of the individual's parents. Panels A and B report estimates for individual annual earnings in levels and percentile ranks, respectively; Panels C and D report estimates for household net wealth in levels and percentile ranks, respectively. PGIs are standardized, so coefficients represent the changes caused by a one-standard-deviation increase in the corresponding PGI. Error bars represent 95% confidence intervals. Panel A and B include 574,690 person-year observations from 38,936 individuals; Panel C and D include 566,810 person-year observations from 38,934 individuals.

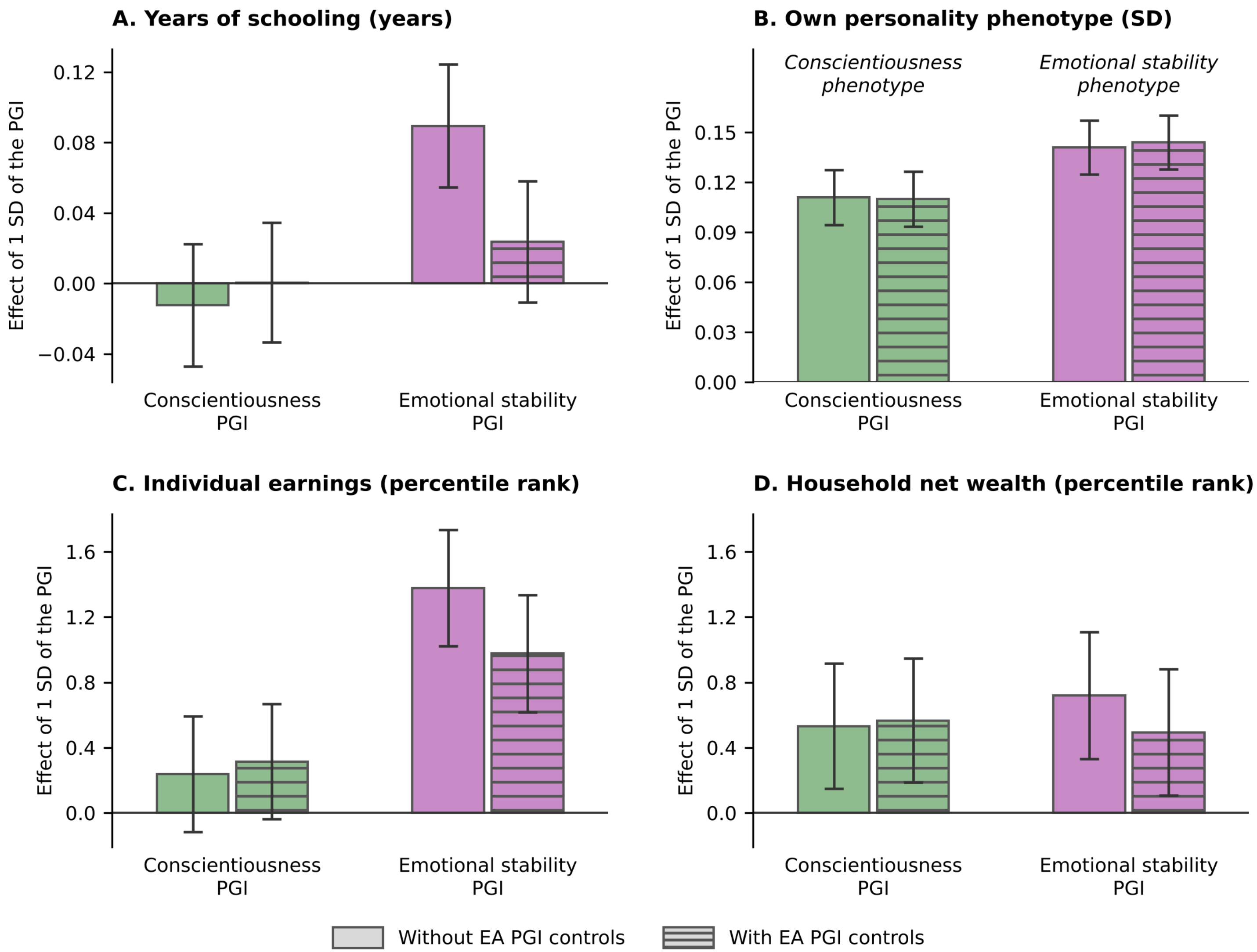

**Fig. 4. Does Genetic Propensity toward Education Account for the Effects of Personality PGIs?** The figure examines how conditioning on genetic propensity toward educational attainment changes the estimated effects of the conscientiousness and emotional stability PGIs, the two Big Five traits for which Figure 3 shows the clearest effects on economic outcomes. Solid bars report estimates from equation (1), while patterned bars report estimates from equation (2), which additionally conditions on the individual's EA PGI and the sum of the EA PGIs of the individual's parents. In each specification, the five coefficients on the personality PGIs are estimated jointly. Estimates for the openness, extraversion, and agreeableness PGIs are reported in Supplementary Materials Table S6. Panels A, B, C, and D report estimates for years of schooling, the individual's own personality phenotype (that is, the effect of the conscientiousness PGI on conscientiousness and of the emotional stability PGI on emotional stability), individual annual earnings, and household net wealth, respectively. PGIs and personality measures are standardized; schooling is measured in years; and earnings and net wealth are measured in percentile ranks. Error bars represent 95% confidence intervals. Panel A includes 37,845 individuals, while Panel B includes 37,768 individuals. Panel C includes 574,690 person-year observations from 38,936 individuals, and Panel D includes 566,810 person-year observations from 38,934 individuals.

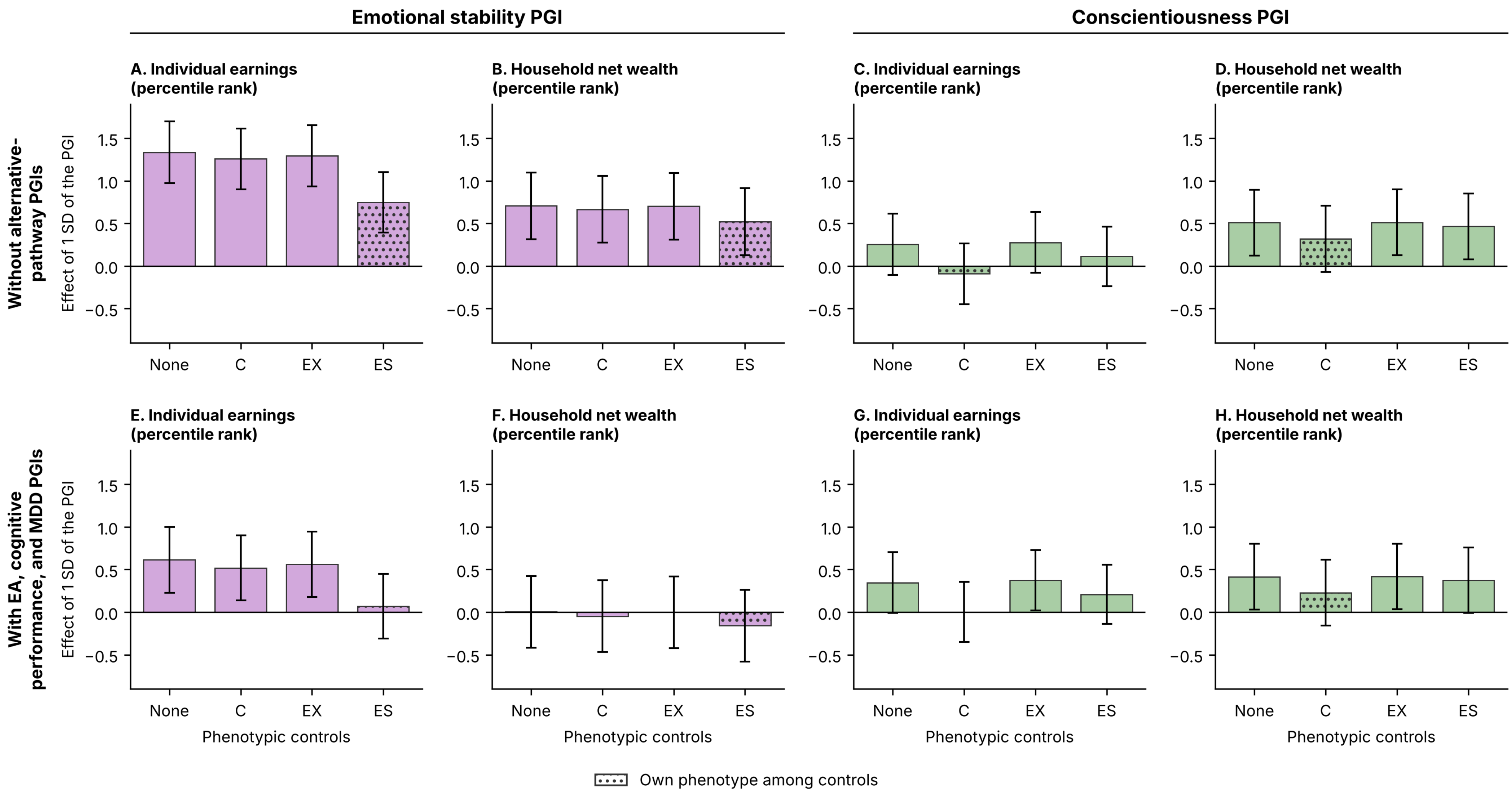

**Fig 5. Descriptive, non-causal mediation analysis.** This figure examines whether the estimated effects of the emotional stability and conscientiousness PGIs on individual earnings and household net wealth operate through the personality traits captured by these PGIs, by adding survey measures of these personality traits to the regressions one at a time. The left two columns report the coefficient on the emotional stability PGI, while the right two columns report the coefficient on the conscientiousness PGI. All regressions simultaneously include the individual's five Big Five personality PGIs and the five Big Five personality PGIs of their parents. Panel titles indicate the dependent variable, which is measured in percentile ranks. Panels in the bottom row additionally control for the educational attainment (EA), cognitive performance, and major depressive disorder (MDD) PGIs of the individual and their parents. Within each panel, estimates correspond to specifications with no phenotypic personality controls or with a survey measure of conscientiousness (C), extraversion (EX), or emotional stability (ES), respectively. Patterned bars identify specifications that control for the survey measure of the same personality trait as the reported PGI. Error bars indicate 95% confidence intervals. Estimates for the effects of the openness, extraversion, and agreeableness PGIs and for the effects on earnings and on wealth measured in euros are shown in Supplementary Materials Figures S1 and S2 and reported in Supplementary Materials Table S10. The earnings panels (A, C, E, and G) include 566,035 person-year observations from 37,765 individuals, and the wealth panels (B, D, F, and H) include 558,346 person-year observations from 37,766 individuals.

Supplementary Materials for

# Random Genetic Variation Links Emotional Stability to Earnings and Wealth

Leandro Carvalho*, Sjoerd van Alten, Silvia Barcellos, Stephen Dorn, Titus Galama, Catharina Hartman, Qiongshi Lu, Dilnoza Muslimova

*Corresponding author. Email: lcarvalh@usc.edu

# 1. Quality Control of genetic data in Lifelines

Lifelines is a multi-disciplinary prospective population-based cohort study examining in a unique three-generation design the health and health-related behaviours of 167,729 persons living in the North of the Netherlands. It employs a broad range of investigative procedures in assessing the biomedical, socio-demographic, behavioural, physical and psychological factors which contribute to the health and disease of the general population, with a special focus on multi-morbidity and complex genetics (Scholtens et al., 2014).

Genotypes in Lifelines were measured from blood samples that were taken during the first assessment visit. Genotyped data was made available in two different subsamples: (1) the GWAS CytoSNP cohort, consisting of 64,488 unrelated respondents, with ~300k SNPs measured on the CytoSNP-12v2 array and (2) the UGLI cohort (second release), which consists of 15,400 respondents, with ~690k 0053NPs measured on the Infinium Global Screening Array®(GSA) MultiEthnic Disease Version. Prior to release, both genotyped subsamples were subjected to strict quality control (Liflines 2026, Sanna et al., 2019; University Medical Center Groningen Genetics Lifelines Initiaive, n.d.). For both subsamples, missing SNPs were imputed using Genome of the Netherlands and 1000 Genomes, resulting in a harmonized set of ~40m SNPs.

Here, we describe further quality control procedures that we undertook on these genotyped Lifelines participants, prior to PGI construction. First, we restricted our analyses to the autosomes. Both the CytoSNP and UGLI subsample were first treated completely separately. Within each subsample, we dropped multiallelic SNPs and loci with MAF < 1%, dropping ~33.5 million loci in CytoSNP, ~27.2 million loci in UGLI. Next, we dropped SNPs with INFO score < 0.8 (dropping 744,661 SNPs in CytoSNP and in 333,409 in UGLI). We also removed participants with excess homozygosity (±3SD, dropping 129 participants in CytoSNP and 529 in UGLI). Then we dropped SNPs that exceeded the HWE p-value threshold of $10^{-6}$ (dropping 1,163 SNPs in CytoSNP and 22,549 in UGLI). 1,289 respondents in the CytoSNP cohort were also present in the UGLI cohort and were therefore dropped from the UGLI cohort. After all these QC steps, ~6.8 million SNPs in CytoSNP and ~7.0 SNPs in UGLI remained.

After all these QC steps, which were performed on each subsample separately, we restricted both subsamples to the ~6.4 million SNPs that they both had in common. We then combined both data sets using the *bmerge* command in PLINK 1.9.

# 2. Polgyenic Index construction

Polygenic indices for each of the Big Five personality traits were constructed using PUMAS-ensemble (Zhao et al., 2024) from GWAS summary statistics that excluded

Lifelines participants (Schwaba et al., 2026). For each trait, we took the intersection of variants present in the GWAS summary statistics, a European-ancestry 1000 Genomes reference panel restricted to HapMap3+ variants (Privé et al., 2020), and the Lifelines genotype data. The 1000 Genomes reference panel was used for linkage disequilibrium (LD) estimation throughout model fitting and summary-statistic subsampling.

Using the PUMAS-ensemble subsampling procedure, we partitioned the original GWAS summary statistics into statistically independent training (70%) and ensemble-learning (30%) sets, repeating this procedure across four cross-validation folds. Within each fold, we fit 108 candidate models in the training set, spanning four methods (lassosum, LDpred2, MegaPRS, and SBayesRC) and a range of tuning parameters (Zhao et al., 2024). PUMAS-ensemble then combined the candidate PGIs using a summary-statistics-based elastic-net regression fit in the ensemble-learning partition. The resulting SNP weights were averaged across the four folds and used to compute the PGIs in Lifelines.

# 3. Mendellian Imputation

To impute parental genotypes, we use SNIPAR ("Single Nucleotide Imputation of PARents", Young et al., 2022), a Python package that imputes missing parental genotypes based on pedigree information. We subsequently create the respective parental polygenic indices using PUMAS based GWAS weights (as explained in Section 2 of the Supplementary Materials) and pgs.py pipeline of the SNIPAR package. The PGIs are based on 646,257 - 914,672 HapMap3 SNPs and 43,392 Lifelines participants with 1st degree relatedness, i.e., parent-offspring pairs and sibling pairs. The final SNP count for each PGI varies depending on the match between the Lifelines sample, imputed SNPs of parents, and the GWAS for the respective trait. We use the same quality control process on the imputation sample as explained in Section 1 of the Supplementary Materials. In addition to that, we apply SNIPAR's protocol to filter out genetic variants with MAF < 5%, and individuals with > 5% missing genotype.

# 4. Coding of earnings and wealth variables

PGIs for Lifelines respondents and the imputed PGIs of their parents were linked to Statistics Netherlands. Statistics Netherlands houses detailed data from tax records on earnings and wealth for the full Dutch population. We used the annual files from 2006 through 2024, since those were the first and last years for which wealth was available. Wealth was measured on January 1st of the reporting year. Earnings cover the whole calendar year and include wages and self-employment profit. Outcomes were available in euros and were expressed in 2021 price levels using CBS's Harmonised Index of Consumer Prices. Net wealth was constructed by CBS as the total assets of a household minus total debt. Percentile rank of both individual earnings and household net wealth was measured by ranking every individual in the full Dutch population within a calendar year and birth-year stratum.

**Fig S1. Descriptive, non-causal mediation analysis: Outcomes in euros.** This figure examines whether the estimated effects of the emotional stability and conscientiousness PGIs on individual earnings and household net wealth operate through the personality traits captured by these PGIs, by adding survey measures of these personality traits to the regressions one at a time. The left two columns report the coefficient on the emotional stability PGI, while the right two columns report the coefficient on the conscientiousness PGI. All regressions simultaneously include the proband's five Big Five personality PGIs and the five Big Five personality PGIs of their parents. Panel titles indicate the dependent variable, which is measured in euros. Panels in the bottom row additionally control for the educational attainment (EA), cognitive performance, and major depressive disorder (MDD) PGIs of the proband and their parents. Within each panel, estimates correspond to specifications with no phenotypic personality controls or with a survey measure of conscientiousness (C), extraversion (EX), or emotional stability (ES), respectively. Patterned bars identify specifications that control for the survey measure of the same personality trait as the reported PGI. Error bars indicate 95% confidence intervals. The figure repeats Figure 5 with outcomes in euros. Estimates are reported in Supplementary Materials Table S10.

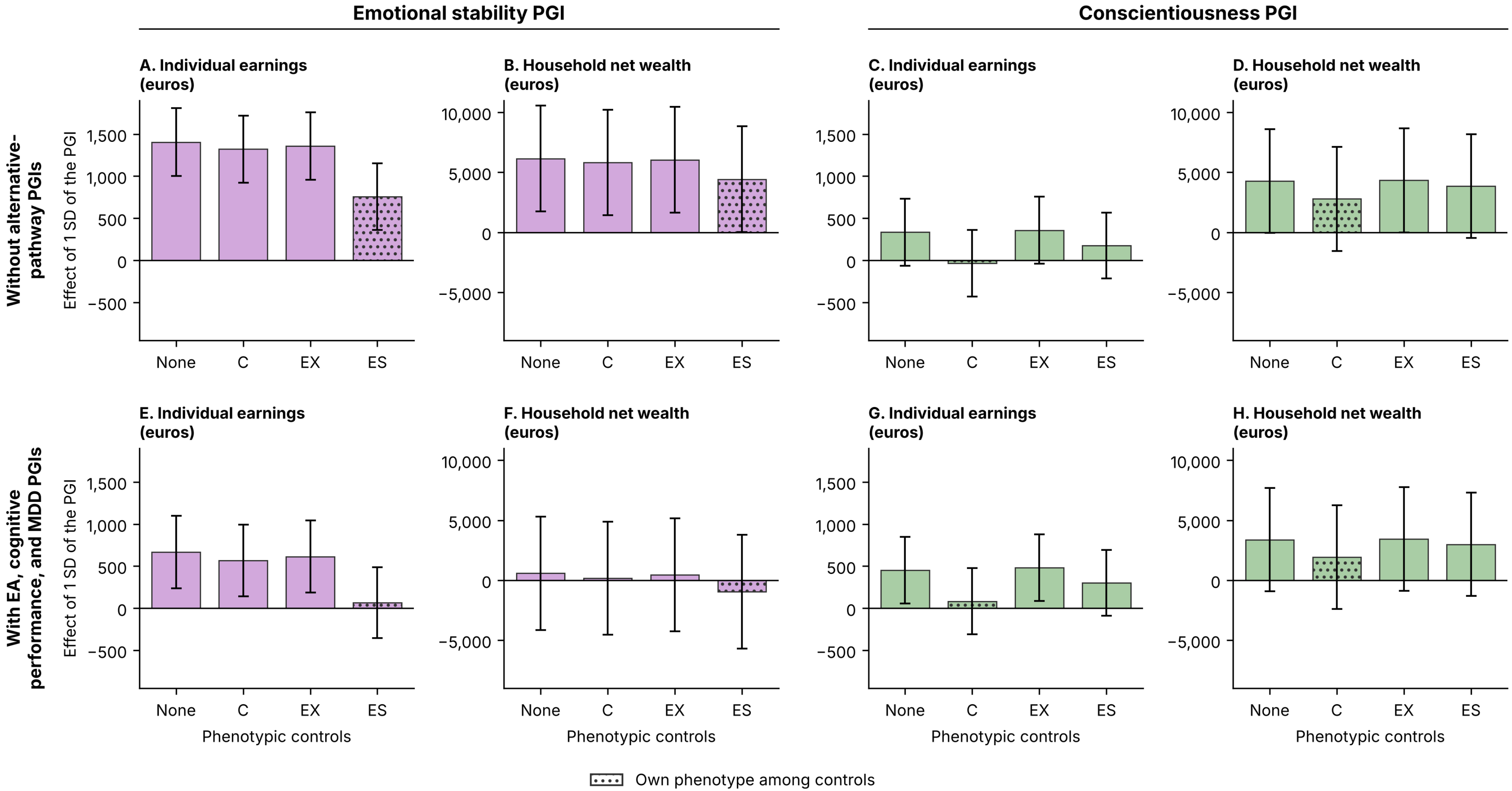

**Fig S2. Descriptive, non-causal mediation analysis: Other Big 5 PGIs.** These figures examine whether the estimated effects of the extraversion, openness, and agreeableness PGIs on individual earnings and household net wealth operate through the personality traits captured by these PGIs, by adding survey measures of these personality traits to the regressions one at a time. Each graph reports the coefficient on the PGI for the indicated personality trait. All regressions simultaneously include the proband's five Big Five personality PGIs and the five Big Five personality PGIs of their parents. Columns correspond to different dependent variables. Panels in the bottom row additionally control for the educational attainment (EA), cognitive performance, and major depressive disorder (MDD) PGIs of the proband and their parents. Within each panel, estimates correspond to specifications with no phenotypic personality controls or with a survey measure of conscientiousness (C), extraversion (EX), or emotional stability (ES), respectively. Patterned bars identify specifications that control for the survey measure of the same personality trait as the reported PGI. No survey measures of openness or agreeableness are available, so Panels I to X have no patterned bars. Error bars indicate 95% confidence intervals. Estimates are reported in Supplementary Materials Table S10.

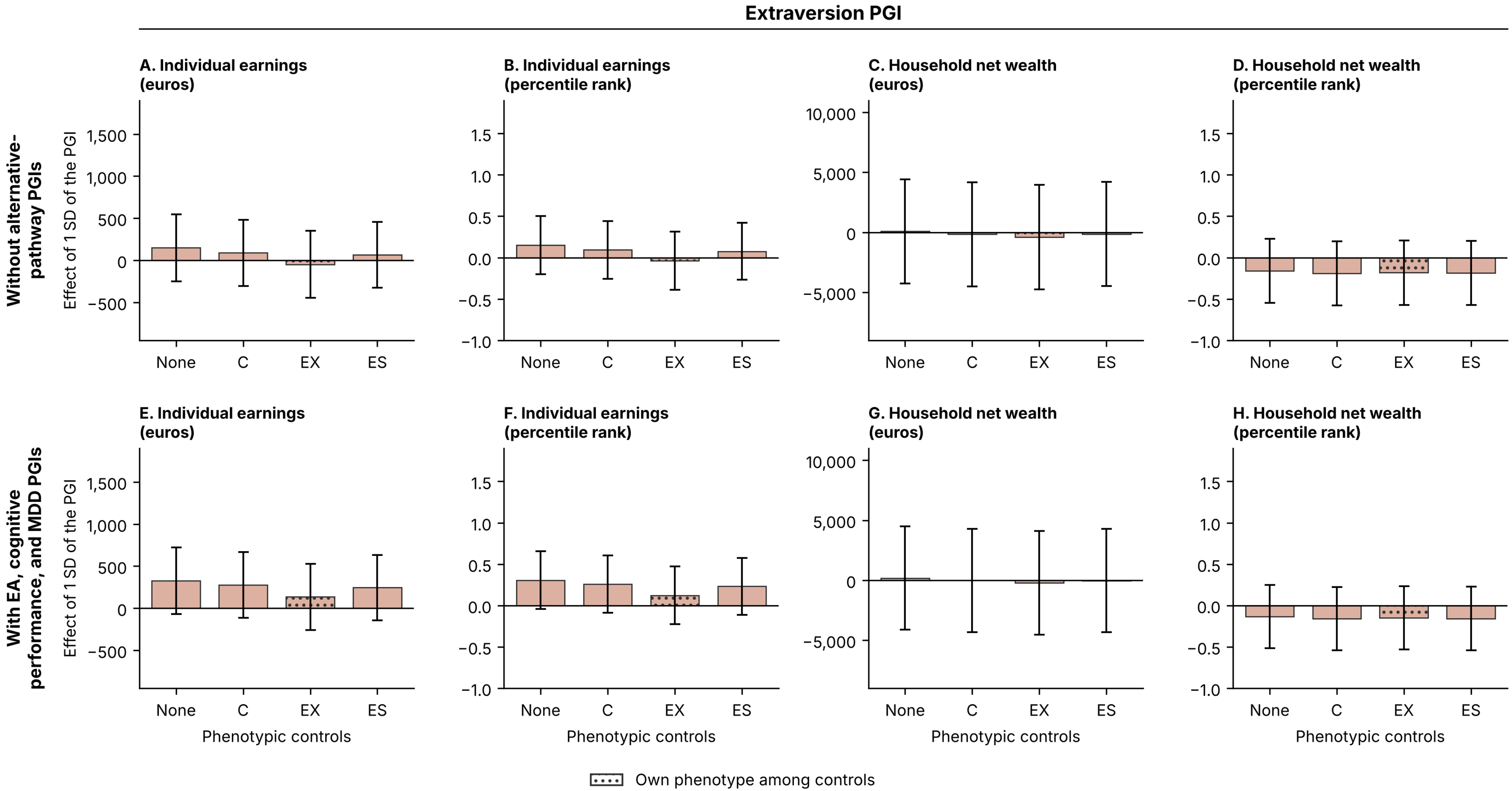

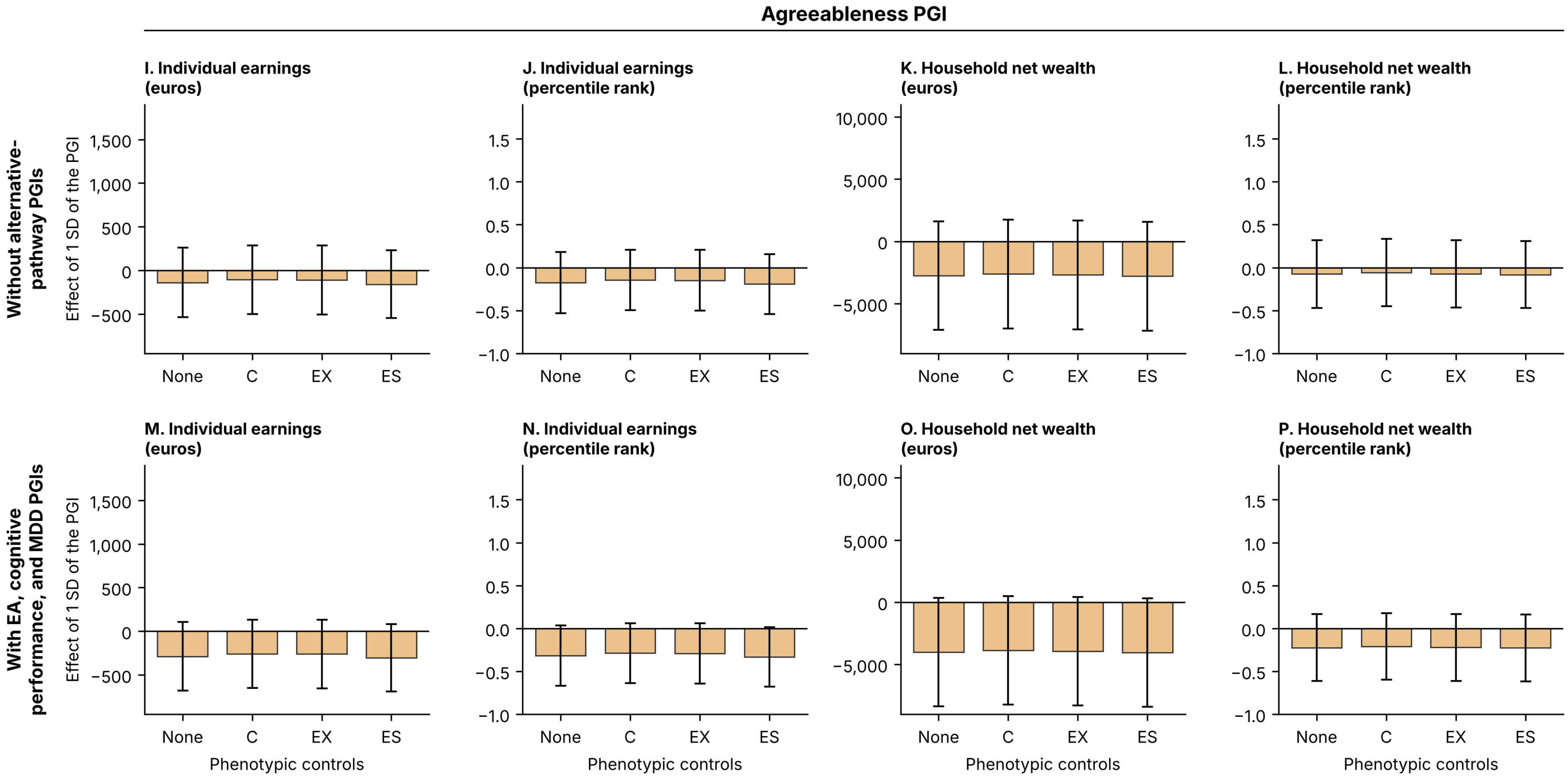

Agreeableness PGI
I. Individual earnings (euros)
J. Individual earnings (percentile rank)
K. Household net wealth (euros)
L. Household net wealth (percentile rank)
M. Individual earnings (euros)
N. Individual earnings (percentile rank)
O. Household net wealth (euros)
P. Household net wealth (percentile rank)
Without alternative-pathway PGIs
With EA, cognitive performance, and MDD PGIs
Effect of 1 SD of the PGI
None
C
EX
ES
Phenotypic controls

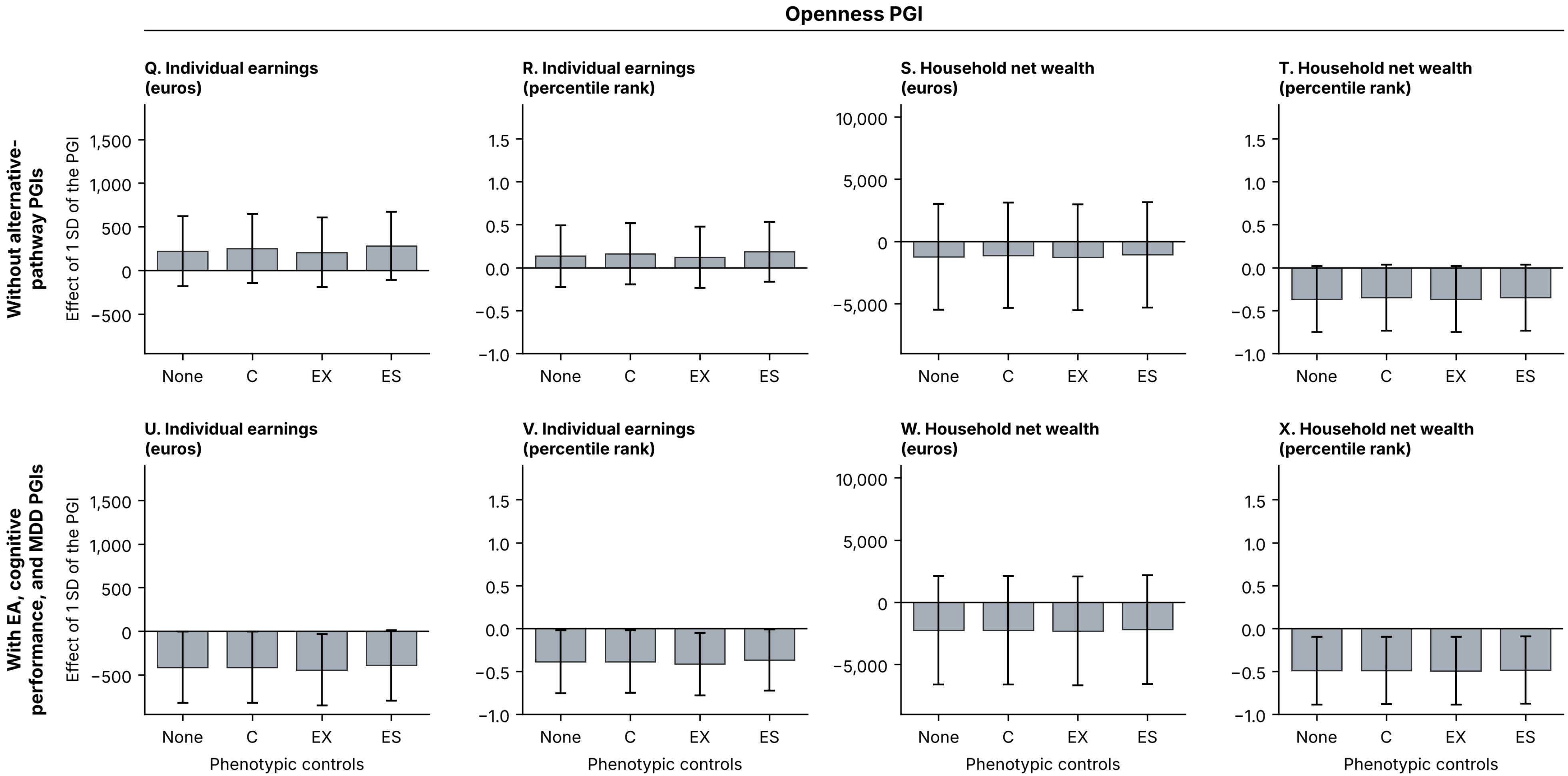

Openness PGI
Q. Individual earnings (euros)
R. Individual earnings (percentile rank)
S. Household net wealth (euros)
T. Household net wealth (percentile rank)
U. Individual earnings (euros)
V. Individual earnings (percentile rank)
W. Household net wealth (euros)
X. Household net wealth (percentile rank)
Without alternative-pathway PGIs
With EA, cognitive performance, and MDD PGIs
Effect of 1 SD of the PGI
None
C
EX
ES
Phenotypic controls

Supplementary Tables for Carvalho et al 2026

***Table S1. Sample Selection***

| Step | N |
|---|---|
| All Lifelines Participants | 167781 |
| Genotyped Lifelines Sample | 78078 |
| Have at least one another parent, sibling or child in Lifelines | 56.449 |
| Final sample aged 30-65 after linkage to CBS, depending on the trait analyze | 33584 - 38941 |

*Table S2. The relationship between the Big 5 PGIs and parental outcomes. Results corresponding to the Figure 1 in the manuscript.*

| | Parents' Household Income | | Parents' Financial Assets | | Parents' Housing Assets | | Parents' Net Wealth | | 1 if parents divorced luring individual's childhoo | | 1 if Father smoked | | 1 if Mother smoked | | Proband's Number of Siblings | | Birthweight | |
|---|---|---|---|---|---|---|---|---|---|---|---|---|---|---|---|---|---|---|
| Agreeableness PGI Proband | 0.263 | 0.0156 | -0.161 | -0.0893 | 0.383 | 0.107 | 0.125 | 0.0608 | -0.00183 | 0.000214 | -0.00575 | -0.00260 | -0.0108 | -0.00976 | 0.00999 | -0.000659 | 2.650 | 1.318 |
| | (0.118) | (0.188) | (0.123) | (0.196) | (0.148) | (0.236) | (0.126) | (0.201) | (0.00142) | (0.00220) | (0.00238) | (0.00379) | (0.00251) | (0.00397) | (0.00394) | (0.00619) | (4.443) | (7.082) |
| Conscientiousness PGI Proband | 0.0134 | 0.165 | 0.517 | 0.235 | 0.255 | 0.261 | 0.386 | 0.152 | 0.000734 | 0.000275 | 0.000633 | 0.000277 | -0.00689 | 0.00281 | -0.00972 | 0.00623 | 2.304 | 6.024 |
| | (0.117) | (0.186) | (0.124) | (0.196) | (0.147) | (0.235) | (0.125) | (0.199) | (0.00142) | (0.00219) | (0.00237) | (0.00377) | (0.00249) | (0.00392) | (0.00369) | (0.00615) | (4.425) | (7.018) |
| Extraversion PGI Proband | -0.100 | 0.0451 | -0.573 | -0.268 | -0.394 | -0.253 | -0.623 | -0.314 | 0.00340 | 1.02e-05 | 0.00247 | -0.00166 | 0.0103 | -0.000420 | -0.0119 | -0.0112 | 7.109 | 7.249 |
| | (0.118) | (0.186) | (0.123) | (0.194) | (0.148) | (0.235) | (0.126) | (0.198) | (0.00142) | (0.00219) | (0.00237) | (0.00372) | (0.00249) | (0.00392) | (0.00369) | (0.00617) | (4.414) | (7.016) |
| Emotional Stability PGI Proband | 0,301 | 0,0469 | 1,333 | 0,131 | 1,456 | -0,0271 | 1,559 | 0,0876 | -0,00648 | 0,001 | -0,0124 | 0,00278 | -0,0195 | -0,00242 | 0,016 | -0,00328 | -2,172 | -7,101 |
| | (0.119) | (0.189) | (0.125) | (0.197) | (0.151) | (0.236) | (0.128) | (0.201) | (0.00140) | (0.00222) | (0.00240) | (0.00378) | (0.00251) | (0.00397) | (0.00393) | (0.00621) | (4.465) | (6.977) |
| Openness PGI Proband | 0.627 | -0.355 | -0.188 | 0.0184 | 0.603 | -0.155 | 0.160 | 0.0324 | 0.00352 | -0.000695 | 0.00299 | 0.00864 | -0.000888 | 0.00718 | 0.00316 | -0.000447 | -3.946 | 0.685 |
| | (0.116) | (0.186) | (0.122) | (0.195) | (0.147) | (0.236) | (0.125) | (0.198) | (0.00139) | (0.00212) | (0.00234) | (0.00371) | (0.00246) | (0.00393) | (0.00371) | (0.00610) | (4.371) | (7.056) |
| Agreeableness PGI Parents | | 0.316 | | -0.0909 | | 0.352 | | 0.0830 | | -0.00262 | | -0.00406 | | -0.00133 | | 0.0139 | | 1.720 |
| | | (0.188) | | (0.196) | | (0.234) | | (0.200) | | (0.00216) | | (0.00380) | | (0.00395) | | (0.00712) | | (7.053) |
| Conscientiousness PGI Parents | | -0.193 | | 0.357 | | -0.00978 | | 0.296 | | 0.000639 | | 0.000554 | | -0.0124 | | -0.0209 | | -4.886 |
| | | (0.185) | | (0.195) | | (0.233) | | (0.198) | | (0.00219) | | (0.00377) | | (0.00393) | | (0.00666) | | (6.939) |
| Extraversion PGI Parents | | -0.179 | | -0.389 | | -0.169 | | -0.390 | | 0.00437 | | 0.00524 | | 0.0138 | | -0.000693 | | -0.185 |
| | | (0.184) | | (0.193) | | (0.232) | | (0.197) | | (0.00218) | | (0.00373) | | (0.00391) | | (0.00667) | | (6.971) |
| Emotional Stability PGI Parents | | -0.320 | | -1.550 | | -1.904 | | -1.894 | | 0.00972 | | 0.0195 | | 0.0219 | | -0.0254 | | -6.447 |
| | | (0.187) | | (0.196) | | (0.235) | | (0.201) | | (0.00221) | | (0.00377) | | (0.00396) | | (0.00677) | | (6.914) |
| Openness PGI Parents | | 1.254 | | -0.266 | | 0.968 | | 0.160 | | 0.00547 | | -0.00714 | | -0.0101 | | 0.00461 | | -5.894 |
| | | (0.185) | | (0.194) | | (0.232) | | (0.197) | | (0.00210) | | (0.00373) | | (0.00391) | | (0.00652) | | (6.898) |
| Constant | 26.49 | 26.61 | 45.73 | 45.85 | 33.16 | 33.40 | 55.59 | 55.79 | 0.00548 | 0.00550 | 0.872 | 0.870 | 0.0868 | 0.0837 | 2.419 | 2.420 | 3,610 | 3,610 |
| | (6.348) | (6.379) | (5.137) | (5.063) | (9.268) | (9.221) | (5.763) | (5.707) | (0.00197) | (0.00250) | (0.0512) | (0.0515) | (0.0386) | (0.0381) | (0.136) | (0.137) | (247.8) | (248.1) |
| Observations | 36,823 | 36,823 | 36,011 | 36,011 | 36,011 | 36,011 | 36,011 | 36,011 | 37,904 | 37,904 | 37,410 | 37,410 | 37,447 | 37,447 | 30,367 | 30,367 | 29,339 | 29,339 |
| R-squared | 0.092 | 0.094 | 0.021 | 0.023 | 0.087 | 0.090 | 0.026 | 0.029 | 0.020 | 0.021 | 0.128 | 0.128 | 0.035 | 0.036 | 0.018 | 0.019 | 0.014 | 0.014 |

***Table S3. Descriptive Statistics of the Analysis Sample.***

| **variable** | **N** | **mean** | **median** | **sd** |
|---|---|---|---|---|
| *Time-invariant characteristics* | | | | |
| Male | 38.941 | 0,38999999 | 0 | 0,49000001 |
| Birthyear (CBS) | 38.941 | 1971,87 | 1971 | 11,8400002 |
| | | | | |
| *Single-measure characteristics* | | | | |
| Years of Schooling | 37.845 | 13,4099998 | 13 | 2,25999999 |
| Cognition | 33.584 | 0,15000001 | 0,239999995 | 0,94999999 |
| Conscientiousness_ | 37.798 | -0,04 | -0,029999999 | 1,01999998 |
| Extraversion_st | 37.768 | 0,06 | -0,07 | 0,99000001 |
| Neuroticism_st | 37.799 | 0,02 | -0,07 | 1,00999999 |
| Any Depression | 37.962 | 0,03 | 0 | 0,17 |
| | | | | |
| *Annual outcomes* | | | | |
| Age | 575.117 | 45,4 | 45 | 9,1 |
| Earnings, levels | 574.690 | € 43.385 | € 41.166 | € 29.586 |
| Net wealth, levels | 566.810 | € 195.543 | € 112.648 | € 297.649 |

*Table S4. Phenotypic associations of personality traits and socioeconomic outcomes in the analysis sample.*

| | Years of Schooling | | | | Annual Individual Earnings *In Levels* | | | | Annual Individual Earnings *In Percentile Ranks* | | | | Household Net Wealth *In Levels* | | | | Household Net Wealth *In Percentile Ranks* | | | |
|---|---|---|---|---|---|---|---|---|---|---|---|---|---|---|---|---|---|---|---|---|
| Phenotypic measure of Conscientiousness | -0,08 | 0.0461 | | | € 3.296 | | | € 1.556 | 3.077 | | | 1.504 | € 14.023 | | | € 9.668 | 1.772 | | | 1.390 |
| | (0.0123) | (0.0116) | | | (126.8) | | | (134.9) | (0.117) | | | (0.124) | (1,352) | | | (1,486) | (0.123) | | | (0.134) |
| Phenotypic measure of Extraversion | 0.00325 | | 0.0522 | | | € 2.244 | | € 1.405 | | 2.157 | | 1.394 | | € 5.830 | | € 2.996 | | 0.263 | | -0.0680 |
| | (0.0124) | | (0.0123) | | | (130.0) | | (129.1) | | (0.118) | | (0.117) | | (1,402) | | (1,429) | | (0.127) | | (0.128) |
| Phenotypic measure of Emotional Stability | -0.341 | | | -0.309 | | | € 4.844 | € 3.987 | | | 4,40 | 3,56 | | | € 14.509 | € 10.156 | | | 1,54 | 0,99 |
| | (0.0129) | | | (0.0119) | | | (125.8) | (136.7) | | | (0.116) | (0.125) | | | (1,326) | (1,477) | | | (0.125) | (0.137) |
| Constant | 9.646 | 9.722 | 9.752 | 9.652 | -€ 2.139 | -€ 880 | -€ 3.292 | -€ 2.263 | 12.62 | 13.83 | 11.57 | 12.59 | € 311.789 | € 315.016 | € 308.274 | € 311.130 | 58.15 | 58.29 | 57.78 | 57.88 |
| | (0.449) | (0.443) | (0.456) | (0.438) | (2,381) | (2,720) | (3,036) | (2,651) | (8.116) | (7.767) | (8.685) | (8.253) | (84,255) | (83,422) | (84,492) | (83,975) | (5.982) | (5.864) | (5.920) | (5.992) |
| Observations | 37,444 | 37,472 | 37,444 | 37,473 | 566,459 | 566,035 | 566,478 | 566,035 | 566,459 | 566,035 | 566,478 | 566,035 | 558,762 | 558,346 | 558,781 | 558,346 | 558,762 | 558,346 | 558,781 | 558,346 |
| R-squared | 0.149 | 0.132 | 0.133 | 0.148 | 0.219 | 0.213 | 0.233 | 0.238 | 0.201 | 0.194 | 0.214 | 0.220 | 0.090 | 0.088 | 0.090 | 0.091 | 0.007 | 0.003 | 0.006 | 0.008 |
| N clusters | 37444 | 37472 | 37444 | 37473 | 37795 | 37765 | 37796 | 37765 | 37795 | 37765 | 37796 | 37765 | 37796 | 37766 | 37797 | 37766 | 37796 | 37766 | 37797 | 37766 |

***Table S5. Summary of polygenic indices used in the analyses.***

| **Trait** | **GWAS Summary Statistics** | **N_GWAS** | Notes |
|---|---|---|---|
| Openness | Schwaba et al (2026) | 509.264 | |
| Conscientiousness | Schwaba et al (2026) | 496.495 | |
| Extraversion | Schwaba et al (2026) | 529.879 | |
| Agreeableness | Schwaba et al (2026) | 505.363 | |
| Emotional Stability | Schwaba et al (2026) | 999.334 | |
| Educational Attainment (EA) | Okbay et al., (2022) | 765.283 | Publically available set (excluding 23andMe) |
| Major Depressive Disorder (MDD) | Adams et al., (2025) | 688.808 | |
| Cognitive Performance (CP) | Lee et al., (2018) | 257.828 | Publically available set (excluding 23andMe) |

***Table S6. Results corresponding to Figures 2-4, including analyses of potential pleiotropy.***

| | Conscientiousness | | | | Extraversion | | | | Emotional Stability | | | |
|---|---|---|---|---|---|---|---|---|---|---|---|---|
| Agreeableness PGI Proband | -0.00894 | -0.00773 | -0.00816 | -0.00861 | -0.00956 | -0.00873 | -0.00902 | -0.00923 | 0,00519 | 0,0059 | 0,00598 | 0,00419 |
| | (0.00839) | (0.00839) | (0.00839) | (0.00842) | (0.00811) | (0.00811) | (0.00812) | (0.00812) | (0.00816) | (0.00817) | (0.00816) | (0.00817) |
| Conscientiousness PGI Proband | 0.111 | 0.110 | 0.108 | 0.107 | -0.0133 | -0.0141 | -0.0156 | -0.0160 | 0,0298 | 0,0291 | 0,0298 | 0,0282 |
| | (0.00836) | (0.00836) | (0.00840) | (0.00840) | (0.00804) | (0.00804) | (0.00808) | (0.00808) | (0.00802) | (0.00802) | (0.00807) | (0.00807) |
| Extraversion PGI Proband | 0.0204 | 0.0189 | 0.0172 | 0.0173 | 0.0882 | 0.0871 | 0.0859 | 0.0858 | 0,017 | 0,0161 | 0,0165 | 0,0169 |
| | (0.00837) | (0.00838) | (0.00840) | (0.00840) | (0.00800) | (0.00800) | (0.00801) | (0.00801) | (0.00816) | (0.00817) | (0.00819) | (0.00818) |
| Emotional Stability PGI Proband | 0,0283 | 0,033 | 0,0342 | 0,0312 | 0,0211 | 0,0247 | 0,0255 | 0,0245 | 0,141 | 0,144 | 0,143 | 0,131 |
| | (0.00857) | (0.00868) | (0.00869) | (0.00928) | (0.00813) | (0.00824) | (0.00826) | (0.00888) | (0.00823) | (0.00832) | (0.00834) | (0.00893) |
| Openness PGI Proband | -0.00600 | 0.000744 | 0.00138 | 0.00246 | 0.00546 | 0.0113 | 0.0117 | 0.0121 | -0,0115 | -0,00683 | -0,007 | -0,00256 |
| | (0.00838) | (0.00864) | (0.00863) | (0.00869) | (0.00801) | (0.00826) | (0.00826) | (0.00834) | (0.00803) | (0.00828) | (0.00828) | (0.00835) |
| Agreeableness PGI Parents | -0.0129 | -0.0130 | -0.0129 | -0.0137 | -0.0167 | -0.0175 | -0.0175 | -0.0189 | 0,0000233 | -0,000546 | -0,00058 | -0,00121 |
| | (0.00832) | (0.00833) | (0.00832) | (0.00835) | (0.00807) | (0.00807) | (0.00807) | (0.00808) | (0.00811) | (0.00812) | (0.00812) | (0.00812) |
| Conscientiousness PGI Parents | 0.0113 | 0.0114 | 0.0108 | 0.0105 | 0.0173 | 0.0180 | 0.0177 | 0.0169 | 0,00885 | 0,00937 | 0,00897 | 0,0088 |
| | (0.00830) | (0.00830) | (0.00834) | (0.00834) | (0.00805) | (0.00806) | (0.00810) | (0.00810) | (0.00807) | (0.00807) | (0.00811) | (0.00811) |
| Extraversion PGI Parents | -0.00246 | -0.00285 | -0.00287 | -0.00252 | -0.0134 | -0.0122 | -0.0122 | -0.0116 | -0,00165 | -0,000909 | -0,00116 | -0,00073 |
| | (0.00843) | (0.00844) | (0.00845) | (0.00845) | (0.00802) | (0.00802) | (0.00803) | (0.00803) | (0.00816) | (0.00817) | (0.00818) | (0.00817) |
| Emotional Stability PGI Parents | -0,0044 | -0,00372 | -0,00342 | -0,00923 | 0,0091 | 0,0053 | 0,0055 | -0,00533 | 0,00597 | 0,00358 | 0,00377 | -0,00261 |
| | (0.00852) | (0.00865) | (0.00866) | (0.00921) | (0.00808) | (0.00820) | (0.00820) | (0.00882) | (0.00817) | (0.00827) | (0.00828) | (0.00888) |
| Openness PGI Parents | -0.00165 | -0.000732 | -0.000519 | 0.00183 | -0.00530 | -0.0114 | -0.0112 | -0.00698 | 0,0105 | 0,00661 | 0,00668 | 0,00947 |
| | (0.00844) | (0.00870) | (0.00870) | (0.00878) | (0.00799) | (0.00824) | (0.00824) | (0.00833) | (0.00804) | (0.00831) | (0.00830) | (0.00838) |
| EA PGI Proband | | -0.0298 | -0.0165 | -0.0177 | | -0.0243 | -0.0153 | -0.0158 | | -0,02 | -0,024 | -0,0288 |
| | | (0.00894) | (0.0101) | (0.0102) | | (0.00855) | (0.00976) | (0.00984) | | (0.00870) | (0.00991) | (0.00997) |
| EA PGI Parents | | -0.00117 | 0.00363 | 0.000316 | | 0.0255 | 0.0289 | 0.0232 | | 0,0165 | 0,0189 | 0,0144 |
| | | (0.00895) | (0.0102) | (0.0104) | | (0.00857) | (0.00980) | (0.00988) | | (0.00870) | (0.0100) | (0.0101) |
| Cog. Performance PGI Proband | | | -0.0265 | -0.0267 | | | -0.0180 | -0.0180 | | | 0,00797 | 0,00727 |
| | | | (0.00965) | (0.00965) | | | (0.00936) | (0.00936) | | | (0.00939) | (0.00940) |
| Cog. Performance PGI Parents | | | -0.00969 | -0.00975 | | | -0.00681 | -0.00704 | | | -0,00461 | -0,00444 |
| | | | (0.00977) | (0.00977) | | | (0.00937) | (0.00937) | | | (0.00947) | (0.00947) |
| MDD PGI Proband | | | | -0.00788 | | | | -0.00243 | | | | -0,0333 |
| | | | | (0.00915) | | | | (0.00885) | | | | (0.00876) |
| MDD PGI Parents | | | | -0.0161 | | | | -0.0300 | | | | -0,0177 |
| | | | | (0.00911) | | | | (0.00880) | | | | (0.00870) |
| Constant | 0.00769 | -0.00637 | -0.0139 | -0.0179 | -0.600 | -0.594 | -0.599 | -0.606 | 0,21 | 0,212 | 0,213 | 0,207 |
| | (0.248) | (0.251) | (0.254) | (0.255) | (0.329) | (0.327) | (0.327) | (0.329) | (0.271) | (0.273) | (0.273) | (0.273) |
| Observations | 37,768 | 37,768 | 37,768 | 37,768 | 37,768 | 37,768 | 37,768 | 37,768 | 37,768 | 37,768 | 37,768 | 37,768 |
| R-squared | 0.036 | 0.037 | 0.038 | 0.038 | 0.058 | 0.059 | 0.059 | 0.060 | 0.062 | 0.062 | 0.062 | 0.064 |
| N clusters | 0.0157 | 0.0166 | 0.0174 | 0.0178 | 0.00797 | 0.00820 | 0.00862 | 0.00943 | 0.0250 | 0.0252 | 0.0252 | 0.0270 |

| | Years of Schooling | | | | Score in Cognitive Test | | | | Any Depression | | | |
|---|---|---|---|---|---|---|---|---|---|---|---|---|
| Agreeableness PGI Proband | 0.0283 | 0.00998 | 0.0115 | 0.0120 | 0.00170 | -0.00179 | -0.000695 | -0.000718 | 0.000941 | 0.00116 | 0.00113 | 0.00150 |
| | (0.0182) | (0.0176) | (0.0176) | (0.0176) | (0.00830) | (0.00828) | (0.00825) | (0.00827) | (0.00142) | (0.00142) | (0.00142) | (0.00142) |
| Conscientiousness PGI Proband | -0.0123 | 0.000504 | 0.0107 | 0.0104 | -0.0456 | -0.0436 | -0.0365 | -0.0365 | 0.00111 | 0.000949 | 0.000530 | 0.000874 |
| | (0.0178) | (0.0173) | (0.0173) | (0.0173) | (0.00803) | (0.00800) | (0.00802) | (0.00803) | (0.00138) | (0.00138) | (0.00139) | (0.00139) |
| Extraversion PGI Proband | -0.0473 | -0.0261 | -0.0187 | -0.0193 | -0.0153 | -0.0121 | -0.00707 | -0.00704 | 0.000435 | 0.000199 | -5.52e-05 | -0.000123 |
| | (0.0177) | (0.0173) | (0.0173) | (0.0173) | (0.00794) | (0.00793) | (0.00791) | (0.00791) | (0.00135) | (0.00135) | (0.00135) | (0.00135) |
| Emotional Stability PGI Proband | 0,0894 | 0,0237 | 0,0184 | 0,0218 | 0,043 | 0,0327 | 0,0287 | 0,0286 | -0,00871 | -0,00797 | -0,00776 | -0,00508 |
| | (0.0178) | (0.0175) | (0.0175) | (0.0187) | (0.00797) | (0.00806) | (0.00803) | (0.00870) | (0.00145) | (0.00145) | (0.00146) | (0.00153) |
| Openness PGI Proband | 0.0934 | 0.00159 | -0.00171 | -0.00291 | 0.0402 | 0.0248 | 0.0220 | 0.0220 | 0.000998 | 0.00202 | 0.00213 | 0.00116 |
| | (0.0178) | (0.0178) | (0.0178) | (0.0180) | (0.00806) | (0.00829) | (0.00826) | (0.00836) | (0.00141) | (0.00145) | (0.00145) | (0.00146) |
| Agreeableness PGI Parents | 0.00133 | -0.00230 | -0.00248 | -0.00631 | -0.00658 | -0.00690 | -0.00665 | -0.00653 | -0.00154 | -0.00145 | -0.00143 | -0.00134 |
| | (0.0181) | (0.0175) | (0.0174) | (0.0175) | (0.00815) | (0.00813) | (0.00810) | (0.00811) | (0.00137) | (0.00137) | (0.00137) | (0.00137) |
| Conscientiousness PGI Parents | 0.0313 | 0.0343 | 0.0334 | 0.0314 | 0.0229 | 0.0235 | 0.0233 | 0.0234 | -0.00125 | -0.00131 | -0.000977 | -0.000979 |
| | (0.0178) | (0.0173) | (0.0173) | (0.0173) | (0.00798) | (0.00795) | (0.00798) | (0.00799) | (0.00138) | (0.00138) | (0.00138) | (0.00138) |
| Extraversion PGI Parents | -0.0104 | 0.00224 | 0.000691 | 0.00239 | 0.00943 | 0.0113 | 0.0106 | 0.0105 | -0.000302 | -0.000531 | -0.000336 | -0.000418 |
| | (0.0176) | (0.0172) | (0.0172) | (0.0172) | (0.00798) | (0.00796) | (0.00793) | (0.00793) | (0.00135) | (0.00136) | (0.00136) | (0.00136) |
| Emotional Stability PGI Parents | 0,0359 | 0,002 | 0,00233 | -0,0262 | -0,00418 | -0,00859 | -0,00847 | -0,0077 | -0,000657 | -0,0000261 | -0,00019 | 0,000717 |
| | (0.0176) | (0.0173) | (0.0173) | (0.0186) | (0.00801) | (0.00808) | (0.00806) | (0.00870) | (0.00142) | (0.00143) | (0.00144) | (0.00152) |
| Openness PGI Parents | 0.0773 | 0.0261 | 0.0261 | 0.0371 | 0.00347 | -0.00283 | -0.00238 | -0.00268 | 0.000422 | 0.00137 | 0.00130 | 0.000879 |
| | (0.0176) | (0.0177) | (0.0177) | (0.0178) | (0.00805) | (0.00824) | (0.00821) | (0.00831) | (0.00138) | (0.00142) | (0.00142) | (0.00143) |
| EA PGI Proband | | 0.417 | 0.356 | 0.357 | | 0.0682 | 0.0263 | 0.0263 | | -0.00474 | -0.00221 | -0.00117 |
| | | (0.0184) | (0.0210) | (0.0212) | | (0.00863) | (0.00971) | (0.00977) | | (0.00153) | (0.00171) | (0.00172) |
| EA PGI Parents | | 0.162 | 0.164 | 0.149 | | 0.0206 | 0.0172 | 0.0176 | | -0.00323 | -0.00529 | -0.00455 |
| | | (0.0185) | (0.0211) | (0.0213) | | (0.00855) | (0.00972) | (0.00983) | | (0.00155) | (0.00173) | (0.00174) |
| Cog. Performance PGI Proband | | | 0.121 | 0.121 | | | 0.0841 | 0.0841 | | | -0.00508 | -0.00491 |
| | | | (0.0200) | (0.0200) | | | (0.00925) | (0.00925) | | | (0.00160) | (0.00160) |
| Cog. Performance PGI Parents | | | -0.00188 | -0.00251 | | | 0.00629 | 0.00630 | | | 0.00403 | 0.00398 |
| | | | (0.0201) | (0.0201) | | | (0.00933) | (0.00934) | | | (0.00159) | (0.00159) |
| MDD PGI Proband | | | | 0.0104 | | | | -0.000311 | | | | 0.00730 |
| | | | | (0.0187) | | | | (0.00866) | | | | (0.00155) |
| MDD PGI Parents | | | | -0.0788 | | | | 0.00215 | | | | 0.00250 |
| | | | | (0.0188) | | | | (0.00874) | | | | (0.00150) |
| Constant | 11.77 | 11.67 | 11.67 | 11.67 | -0.781 | -0.802 | -0.803 | -0.803 | 0.0233 | 0.0253 | 0.0255 | 0.0267 |
| | (0.307) | (0.294) | (0.294) | (0.294) | (0.142) | (0.143) | (0.143) | (0.143) | (0.0163) | (0.0163) | (0.0164) | (0.0165) |
| Observations | 37,845 | 37,845 | 37,845 | 37,845 | 33,584 | 33,584 | 33,584 | 33,584 | 37,962 | 37,962 | 37,962 | 37,962 |
| R-squared | 0.110 | 0.164 | 0.166 | 0.167 | 0.071 | 0.078 | 0.084 | 0.084 | 0.006 | 0.007 | 0.008 | 0.010 |
| N clusters | | | | | | | | | | | | |

| | **Annual Individual Earnings** *In Levels* | | | | **Annual Individual Earnings** *In Percentile Ranks* | | | | **Household Net Wealth** *In Levels* | | | | **Household Net Wealth** *In Percentile Ranks* | | | |
|---|---|---|---|---|---|---|---|---|---|---|---|---|---|---|---|---|
| Agreeableness PGI Proband | -115.6 | -248.7 | -239.5 | -268.5 | -0.151 | -0.267 | -0.260 | -0.296 | -3,083 | -3,670 | -3,823 | -4,316 | -0.0929 | -0.162 | -0.176 | -0.239 |
| | (200.9) | (199.2) | (199.1) | (199.2) | (0.180) | (0.179) | (0.179) | (0.179) | (2,213) | (2,203) | (2,202) | (2,205) | (0.199) | (0.197) | (0.197) | (0.197) |
| Conscientiousness PGI Proband | 312.7 | 401.0 | 458.3 | 429.1 | 0.237 | 0.314 | 0.362 | 0.327 | 4,344 | 4,591 | 3,952 | 3,445 | 0.531 | 0.565 | 0.500 | 0.437 |
| | (202.1) | (200.3) | (201.2) | (201.2) | (0.181) | (0.180) | (0.181) | (0.181) | (2,190) | (2,181) | (2,186) | (2,183) | (0.196) | (0.194) | (0.195) | (0.194) |
| Extraversion PGI Proband | 117.0 | 248.6 | 289.3 | 297.5 | 0.104 | 0.219 | 0.253 | 0.264 | 377.4 | 855.3 | 361.7 | 497.3 | -0.155 | -0.0954 | -0.144 | -0.127 |
| | (200.8) | (199.4) | (199.5) | (199.4) | (0.179) | (0.178) | (0.178) | (0.178) | (2,197) | (2,185) | (2,185) | (2,181) | (0.197) | (0.195) | (0.195) | (0.194) |
| Emotional Stability PGI Proband | 1420 | 957,3 | 926,8 | 690,8 | 1,378 | 0,976 | 0,95 | 0,658 | 6322 | 4464 | 4798 | 762 | 0,72 | 0,494 | 0,528 | 0,0203 |
| | (203.2) | (203.7) | (203.9) | (217.2) | (0.182) | (0.183) | (0.183) | (0.196) | (2,219) | (2,225) | (2,227) | (2,382) | (0.198) | (0.198) | (0.198) | (0.213) |
| Openness PGI Proband | 247.1 | -446.2 | -464.8 | -382.9 | 0.166 | -0.436 | -0.452 | -0.350 | -1,318 | -3,888 | -3,655 | -2,257 | -0.364 | -0.683 | -0.660 | -0.484 |
| | (202.4) | (205.3) | (205.4) | (207.3) | (0.181) | (0.184) | (0.184) | (0.185) | (2,140) | (2,194) | (2,195) | (2,209) | (0.194) | (0.198) | (0.198) | (0.200) |
| Agreeableness PGI Parents | 25.25 | 20.22 | 18.62 | -6.246 | 0.113 | 0.107 | 0.105 | 0.0843 | 1,094 | 525.4 | 526.4 | 66.45 | -0.152 | -0.204 | -0.203 | -0.254 |
| | (202.6) | (200.9) | (200.7) | (201.0) | (0.182) | (0.180) | (0.180) | (0.181) | (2,228) | (2,218) | (2,216) | (2,221) | (0.198) | (0.197) | (0.196) | (0.196) |
| Conscientiousness PGI Parents | 37.80 | 55.49 | 43.39 | 40.97 | -0.00353 | 0.0132 | 0.00249 | 0.00317 | 2,013 | 2,610 | 2,371 | 2,320 | 0.290 | 0.346 | 0.332 | 0.328 |
| | (201.2) | (199.0) | (200.1) | (200.0) | (0.181) | (0.179) | (0.180) | (0.180) | (2,202) | (2,188) | (2,189) | (2,189) | (0.196) | (0.194) | (0.195) | (0.195) |
| Extraversion PGI Parents | -164.4 | -80.00 | -95.34 | -89.97 | -0.180 | -0.104 | -0.117 | -0.113 | -1,399 | 9.653 | 14.14 | 119.1 | -0.199 | -0.0624 | -0.0572 | -0.0458 |
| | (201.6) | (200.2) | (200.3) | (200.3) | (0.180) | (0.179) | (0.179) | (0.179) | (2,177) | (2,166) | (2,166) | (2,162) | (0.197) | (0.196) | (0.196) | (0.195) |
| Emotional Stability PGI Parents | -85,28 | -258,8 | -252,8 | -419,5 | -0,111 | -0,269 | -0,264 | -0,407 | 8046 | 4394 | 4526 | 1463 | 0,73 | 0,382 | 0,389 | 0,0465 |
| | (203.7) | (204.3) | (204.5) | (217.6) | (0.182) | (0.182) | (0.183) | (0.195) | (2,173) | (2,185) | (2,187) | (2,339) | (0.197) | (0.198) | (0.198) | (0.212) |
| Openness PGI Parents | 451.7 | 223.9 | 228.4 | 302.9 | 0.323 | 0.113 | 0.117 | 0.184 | 915.4 | -4,471 | -4,447 | -3,088 | -0.0169 | -0.527 | -0.527 | -0.373 |
| | (203.0) | (206.5) | (206.5) | (208.8) | (0.182) | (0.185) | (0.185) | (0.187) | (2,152) | (2,202) | (2,202) | (2,216) | (0.194) | (0.199) | (0.199) | (0.200) |
| EA PGI Proband | | 3,057 | 2,716 | 2,621 | | 2.657 | 2.369 | 2.253 | | 12,070 | 15,701 | 14,080 | | 1.472 | 1.844 | 1.641 |
| | | (211.8) | (241.0) | (242.3) | | (0.189) | (0.215) | (0.217) | | (2,306) | (2,577) | (2,593) | | (0.206) | (0.234) | (0.235) |
| EA PGI Parents | | 677.4 | 682.0 | 675.3 | | 0.635 | 0.684 | 0.580 | | 20,299 | -7,205 | -7,319 | | 1.901 | 2.028 | 1.798 |
| | | (210.2) | (233.7) | (233.6) | | (0.189) | (0.216) | (0.218) | | (2,343) | (2,464) | (2,461) | | (0.207) | (0.236) | (0.237) |
| Cog. Performance PGI Proband | | | 732.2 | -635.7 | | | 0.575 | 0.566 | | | 22,281 | -10,862 | | | -0.739 | -0.755 |
| | | | (240.0) | (214.7) | | | (0.209) | (0.208) | | | (2,643) | (2,399) | | | (0.223) | (0.222) |
| Cog. Performance PGI Parents | | | -105.7 | 621.3 | | | -0.0961 | -0.0950 | | | -3,948 | 20,275 | | | -0.254 | -0.255 |
| | | | (235.4) | (242.1) | | | (0.210) | (0.210) | | | (2,496) | (2,664) | | | (0.225) | (0.224) |
| | | | | | | | [0.647] | [0.651] | | | | | | | [0.259] | [0.256] |
| MDD PGI Proband | | | | -106.5 | | | | -0.789 | | | | -3,966 | | | | -1.368 |
| | | | | (235.3) | | | | (0.192) | | | | (2,491) | | | | (0.210) |
| MDD PGI Parents | | | | -466.8 | | | | -0.403 | | | | -8,578 | | | | -0.961 |
| | | | | (214.6) | | | | (0.193) | | | | (2,344) | | | | (0.210) |
| Constant | -3,360 | -3,896 | -3,917 | -4,044 | 14.46 | 13.98 | 13.96 | 13.83 | 370,438 | 364,104 | 364,483 | 362,301 | 59.60 | 58.96 | 59.00 | 58.74 |
| | (2,846) | (2,840) | (2,839) | (2,842) | (4.574) | (4.523) | (4.521) | (4.518) | (46,622) | (45,546) | (45,556) | (45,499) | (3.107) | (3.051) | (3.046) | (3.056) |
| Observations | 574,690 | 574,690 | 574,690 | 574,690 | 574,690 | 574,690 | 574,690 | 574,690 | 566,810 | 566,810 | 566,810 | 566,810 | 566,810 | 566,810 | 566,810 | 566,810 |
| R-squared | 0.211 | 0.225 | 0.225 | 0.226 | 0.190 | 0.203 | 0.203 | 0.204 | 0.091 | 0.100 | 0.101 | 0.104 | 0.006 | 0.019 | 0.020 | 0.025 |
| N clusters | 38936 | 38936 | 38936 | 38936 | 38936 | 38936 | 38936 | 38936 | 38934 | 38934 | 38934 | 38934 | 38934 | 38934 | 38934 | 38934 |

*Table S7. Attenuation due to inclusion of parental PGIs.*

| | Conscientiousness | | Extraversion | | Emotional Stability | | individual Earnings | | household net wealth | |
|---|---|---|---|---|---|---|---|---|---|---|
| Agreeableness PGI Proband | -0.0190 | -0.00894 | -0.0225 | -0.00956 | 0,0052 | 0,00519 | -€ 95 | -€ 116 | -€ 2.235 | -€ 3.083 |
| | (0.00533) | (0.00839) | (0.00509) | (0.00811) | (0.00514) | (0.00816) | (125.9) | (200.9) | (1,398) | (2,213) |
| Conscientiousness PGI Proband | 0.120 | 0.111 | 0.000212 | -0.0133 | 0,0367 | 0,0298 | € 342 | € 313 | € 5.925 | € 4.344 |
| | (0.00530) | (0.00836) | (0.00511) | (0.00804) | (0.00511) | (0.00802) | (127.2) | (202.1) | (1,425) | (2,190) |
| Extraversion PGI Proband | 0.0185 | 0.0204 | 0.0779 | 0.0882 | 0,0157 | 0,017 | -€ 11 | € 117 | -€ 730 | € 377 |
| | (0.00526) | (0.00837) | (0.00507) | (0.00800) | (0.00517) | (0.00816) | (126.5) | (200.8) | (1,396) | (2,197) |
| Emotional Stability PGI Proband | 0,0249 | 0,0283 | 0,0281 | 0,0211 | 0.145 | 0.141 | € 1.358 | € 1.420 | € 12.524 | € 6.322 |
| | (0.00541) | (0.00857) | (0.00513) | (0.00813) | (0.00523) | (0.00823) | (128.9) | (203.2) | (1,420) | (2,219) |
| Openness PGI Proband | -0.00731 | -0.00600 | 0.00138 | 0.00546 | -0,00333 | -0,0115 | € 598 | € 247 | -€ 573 | -€ 1.318 |
| | (0.00519) | (0.00838) | (0.00498) | (0.00801) | (0.00498) | (0.00803) | (124.9) | (202.4) | (1,355) | (2,140) |
| Agreeableness PGI Parents | | -0.0129 | | -0.0167 | | 2,33E-05 | | 25.25 | | 1,094 |
| | | (0.00832) | | (0.00807) | | (0.00811) | | (202.6) | | (2,228) |
| Conscientiousness PGI Parents | | 0.0113 | | 0.0173 | | 0,00885 | | 37.80 | | 2,013 |
| | | (0.00830) | | (0.00805) | | (0.00807) | | (201.2) | | (2,202) |
| Extraversion PGI Parents | | -0.00246 | | -0.0134 | | -0,00165 | | -164.4 | | -1,399 |
| | | (0.00843) | | (0.00802) | | (0.00816) | | (201.6) | | (2,177) |
| Emotional Stability PGI Parents | | -0,0044 | | 0,0091 | | 0.00597 | | -€ 85 | | € 8.046 |
| | | (0.00852) | | (0.00808) | | (0.00817) | | (203.7) | | (2,173) |
| Openness PGI Parents | | -0.00165 | | -0.00530 | | 0,0105 | | € 452 | | € 915 |
| | | (0.00844) | | (0.00799) | | (0.00804) | | (203.0) | | (2,152) |
| Constant | 0.00899 | 0.00769 | -0.602 | -0.600 | -0.205 | -0.210 | -3.414 | -3.360 | 369,788 | 370,438 |
| | (0.247) | (0.248) | (0.330) | (0.329) | (0.270) | (0.271) | (2,841) | (2,846) | (46,655) | (46,622) |
| Observations | 37,768 | 37,768 | 37,768 | 37,768 | 37,768 | 37,768 | 574,690 | 574,690 | 566,810 | 566,810 |
| R-squared | 0.036 | 0.036 | 0.058 | 0.058 | 0.062 | 0.062 | 0.211 | 0.211 | 0.091 | 0.091 |

***Table S8. Relationships between PGIs used in the analysis.***

| | Agreeableness PGI | Conscientiousness PGI | Extraversion PGI | Emotional Stability PGI | Openness PGI | EA PGI | Cog. Performance PGI | MDD PGI | EA PGI | Cog. Performance PGI | MDD PGI |
|---|---|---|---|---|---|---|---|---|---|---|---|
| Agreeableness PGI Proband | | 0.130 | 0.0527 | 0,149 | 0.0472 | 0.0364 | 0.0109 | -0.0544 | 0.0109 | -0.00384 | -0.0195 |
| | | (0.00534) | (0.00535) | (0.00530) | (0.00543) | (0.00508) | (0.00526) | (0.00500) | (0.00270) | (0.00281) | (0.00310) |
| Conscientiousness PGI Proband | 0.128 | | 0.0934 | 0,0765 | -0.0460 | -0.0333 | -0.0981 | -0.0307 | 0.00177 | -0.0318 | -0.0178 |
| | (0.00523) | | (0.00527) | (0.00529) | (0.00533) | (0.00503) | (0.00528) | (0.00492) | (0.00268) | (0.00279) | (0.00311) |
| Extraversion PGI Proband | 0.0516 | 0.0930 | | 0,15 | 0.108 | -0.0464 | -0.0739 | 0.0263 | -0.00379 | -0.0211 | 0.00698 |
| | (0.00523) | (0.00525) | | (0.00526) | (0.00528) | (0.00503) | (0.00521) | (0.00490) | (0.00270) | (0.00279) | (0.00310) |
| Emotional Stability PGI Proband | 0,149 | 0,0774 | 0,152 | | 0,0082 | 0,151 | 0,118 | -0,386 | 0,0229 | 0,0156 | -0,162 |
| | (0.00524) | (0.00535) | (0.00531) | | (0.00541) | (0.00502) | (0.00526) | (0.00496) | (0.00293) | (0.00308) | (0.00331) |
| Openness PGI Proband | 0.0461 | -0.0457 | 0.108 | 0,00805 | | 0.236 | 0.140 | 0.0939 | 0.0739 | 0.0122 | 0.0587 |
| | (0.00530) | (0.00530) | (0.00528) | (0.00531) | | (0.00501) | (0.00523) | (0.00492) | (0.00275) | (0.00285) | (0.00316) |
| EA PGI Proband | | | | | | | | | | 0.502 | -0.156 |
| | | | | | | | | | | (0.00485) | (0.00611) |
| Cog. Performance PGI Proband | | | | | | | | | 0.463 | | -0.0253 |
| | | | | | | | | | (0.00448) | | (0.00589) |
| MDD PGI Proband | | | | | | | | | -0.116 | -0.0204 | |
| | | | | | | | | | (0.00457) | (0.00476) | |
| Agreeableness PGI Parents | 0.777 | -0.102 | -0.0425 | -0,112 | -0.0366 | -0.0283 | -0.00919 | 0.0451 | | | |
| | (0.00324) | (0.00531) | (0.00534) | (0.00524) | (0.00532) | (0.00499) | (0.00523) | (0.00494) | | | |
| Conscientiousness PGI Parents | -0.0994 | 0.776 | -0.0714 | -0,0585 | 0.0365 | 0.0252 | 0.0776 | 0.0211 | | | |
| | (0.00519) | (0.00323) | (0.00527) | (0.00528) | (0.00527) | (0.00500) | (0.00524) | (0.00488) | | | |
| Extraversion PGI Parents | -0.0408 | -0.0707 | 0.773 | -0,119 | -0.0865 | 0.0359 | 0.0568 | -0.0208 | | | |
| | (0.00520) | (0.00522) | (0.00324) | (0.00522) | (0.00527) | (0.00499) | (0.00520) | (0.00486) | | | |
| Emotional Stability PGI Parents | -0,114 | -0,0623 | -0,122 | 0.774 | -0,00882 | -0,114 | -0,0895 | 0,298 | | | |
| | (0.00527) | (0.00529) | (0.00526) | (0.00327) | (0.00534) | (0.00501) | (0.00524) | (0.00508) | | | |
| Openness PGI Parents | -0.0382 | 0.0330 | -0.0819 | -0,00481 | 0.781 | -0.184 | -0.107 | -0.0724 | | | |
| | (0.00526) | (0.00522) | (0.00521) | (0.00525) | (0.00318) | (0.00499) | (0.00517) | (0.00489) | | | |
| EA PGI Parents | | | | | | 0.794 | | | 0.777 | -0.405 | 0.121 |
| | | | | | | (0.00307) | | | (0.00318) | (0.00500) | (0.00608) |
| Cog. Performance PGI Parents | | | | | | | 0.782 | | -0.363 | 0.780 | 0.0200 |
| | | | | | | | (0.00318) | | (0.00468) | (0.00323) | (0.00587) |
| MDD PGI Parents | | | | | | | | 0.775 | 0.0865 | 0.0165 | 0.718 |
| | | | | | | | | (0.00326) | (0.00439) | (0.00458) | (0.00336) |
| Observations | 38,941 | 38,941 | 38,941 | 38,941 | 38,941 | 38,941 | 38,941 | 38,941 | 38,941 | 38,941 | 38,941 |
| R-squared | 0.625 | 0.618 | 0.616 | 0.622 | 0.615 | 0.663 | 0.631 | 0.670 | 0.734 | 0.712 | 0.643 |

**Table S9. The relationship between alternative pathway PGIs and their corresponding traits.**

| | Years of Education | | | | | Cognitive Performance (Test Score) | | | | | Any Depression Diagnosis | | | | |
|---|---|---|---|---|---|---|---|---|---|---|---|---|---|---|---|
| EA PGI Proband | 0.421 | | | 0.361 | 0.361 | 0.0817 | | | 0.0366 | 0.0361 | -0.00544 | | | -0.00255 | -0.000995 |
| | (0.0177) | | | (0.0206) | (0.0208) | (0.00836) | | | (0.00953) | (0.00962) | (0.00149) | | | (0.00170) | (0.00170) |
| EA PGI Parents | 0.171 | | | 0.174 | 0.159 | 0.0163 | | | 0.0139 | 0.0142 | -0.00326 | | | -0.00567 | -0.00478 |
| | (0.0177) | | | (0.0207) | (0.0209) | (0.00829) | | | (0.00955) | (0.00969) | (0.00150) | | | (0.00170) | (0.00171) |
| Cognitive Performance PGI Proband | | 0.289 | | 0.118 | 0.118 | | 0.105 | | 0.0878 | 0.0878 | | -0.00683 | | -0.00565 | -0.00543 |
| | | (0.0176) | | (0.0200) | (0.0200) | | (0.00815) | | (0.00931) | (0.00931) | | (0.00142) | | (0.00161) | (0.00161) |
| Cognitive Performance PGI Parents | | 0.101 | | -0.00511 | -0.00586 | | 0.0138 | | 0.00484 | 0.00489 | | 0.00151 | | 0.00461 | 0.00455 |
| | | (0.0176) | | (0.0201) | (0.0201) | | (0.00812) | | (0.00938) | (0.00938) | | (0.00142) | | (0.00161) | (0.00161) |
| MDD PGI Proband | | | -0.0714 | | 0.00455 | | | -0.0186 | | -0.00302 | | | 0.00977 | | 0.00902 |
| | | | (0.0175) | | (0.0172) | | | (0.00788) | | (0.00792) | | | (0.00145) | | (0.00146) |
| MDD PGI Parents | | | -0.124 | | -0.0701 | | | -0.00455 | | 0.00160 | | | 0.00323 | | 0.00259 |
| | | | (0.0175) | | (0.0173) | | | (0.00793) | | (0.00803) | | | (0.00140) | | (0.00141) |
| Observations | 36,846 | 36,846 | 36,846 | 36,846 | 36,846 | 32,681 | 32,681 | 32,681 | 32,681 | 32,681 | 36,962 | 36,962 | 36,962 | 36,962 | 36,962 |
| R-squared | 0.163 | 0.128 | 0.107 | 0.165 | 0.166 | 0.068 | 0.073 | 0.059 | 0.075 | 0.075 | 0.005 | 0.004 | 0.008 | 0.005 | 0.009 |
| Increase in R-squared over controls | 0.0632 | 0.0275 | 0.00676 | 0.0651 | 0.0660 | 0.0102 | 0.0152 | 0.000559 | 0.0172 | 0.0172 | 0.00242 | 0.00116 | 0.00549 | 0.00275 | 0.00695 |

***Table S10. Non-causal, descriptive mediation analysis. Results corresponding to Figure 5 and Supplementary Materials Figure 1.***

| | Individual Earnings — Levels (€) | | | | | | | |
|---|---|---|---|---|---|---|---|---|
| Agreeableness PGI Proband | -138.2 | -106.3 | -109.8 | -157.8 | -288.0 | -257.7 | -261.4 | -304.2 |
| | (202.6) | (200.9) | (201.8) | (198.8) | (200.8) | (199.1) | (200.0) | (197.0) |
| Conscientiousness PGI Proband | 334.7 | -35.14 | 357.3 | 177.4 | 453.0 | 83.81 | 481.0 | 301.5 |
| | (203.6) | (202.4) | (202.9) | (199.2) | (202.7) | (201.2) | (202.0) | (198.4) |
| Extraversion PGI Proband | 148.3 | 88.67 | -47.28 | 66.28 | 327.2 | 276.4 | 135.5 | 244.4 |
| | (202.6) | (200.4) | (202.5) | (198.5) | (201.2) | (199.0) | (201.1) | (197.2) |
| Emotional Stability PGI Proband | 1.404 | 1.321 | 1.360 | 756,5 | 668,4 | 567,4 | 613,6 | 68,17 |
| | (205.3) | (203.0) | (204.4) | (201.2) | (219.2) | (216.9) | (218.2) | (215.1) |
| Openness PGI Proband | 221.3 | 250.7 | 206.9 | 279.7 | -412.2 | -413.0 | -442.2 | -391.0 |
| | (204.5) | (203.0) | (203.5) | (199.9) | (209.4) | (207.8) | (208.4) | (205.1) |
| Agreeableness PGI Parents | 41.83 | 86.25 | 76.42 | 33.25 | 7.033 | 57.20 | 47.93 | 5.727 |
| | (204.2) | (202.4) | (203.5) | (200.3) | (202.6) | (200.6) | (201.8) | (198.6) |
| Conscientiousness PGI Parents | 28.35 | -1.071 | -3.927 | -2.242 | 30.64 | 1.764 | 0.0284 | -0.0574 |
| | (202.7) | (201.1) | (201.9) | (198.7) | (201.6) | (199.7) | (200.7) | (197.4) |
| Extraversion PGI Parents | -191.9 | -188.4 | -156.8 | -178.6 | -117.5 | -115.2 | -85.25 | -106.7 |
| | (203.2) | (201.0) | (202.6) | (199.0) | (202.0) | (199.6) | (201.3) | (197.7) |
| Emotional Stability PGI Parents | -96,37 | -84,75 | -121,9 | -141,6 | -425 | -389,2 | -416,7 | -428,8 |
| | (205.5) | (203.2) | (204.5) | (201.1) | (219.5) | (217.1) | (218.5) | (215.0) |
| Openness PGI Parents | 489.8 | 474.6 | 501.8 | 427.3 | 338.4 | 315.2 | 351.4 | 277.6 |
| | (205.2) | (203.5) | (204.3) | (200.4) | (211.0) | (209.3) | (210.1) | (206.3) |
| EA PGI Proband | | | | | 2,629 | 2,702 | 2,676 | 2,756 |
| | | | | | (244.4) | (243.0) | (243.9) | (239.9) |
| EA PGI Parents | | | | | 701.3 | 792.9 | 731.4 | 653.9 |
| | | | | | (235.6) | (233.3) | (235.0) | (230.6) |
| Cog. Performance PGI Proband | | | | | -628.2 | -613.3 | -630.1 | -481.4 |
| | | | | | (216.4) | (214.1) | (215.6) | (212.1) |
| Cog. Performance PGI Parents | | | | | 616.0 | 602.1 | 563.2 | 559.7 |
| | | | | | (244.3) | (242.4) | (243.6) | (240.2) |
| MDD PGI Proband | | | | | -110.9 | -90.04 | -81.12 | -92.58 |
| | | | | | (237.5) | (234.8) | (236.9) | (232.6) |
| MDD PGI Parents | | | | | -465.5 | -399.7 | -389.4 | -376.0 |
| | | | | | (216.4) | (214.4) | (215.7) | (212.1) |
| Conscientiousness Phenotype | | 3,255 | | | | 3,359 | | |
| | | (127.8) | | | | (125.9) | | |
| Extraversion Phenotype | | | 2,200 | | | | 2,216 | |
| | | | (130.2) | | | | (128.6) | |
| Emotional Stability Phenotype | | | | 4.730 | | | | 4.711 |
| | | | | (127.4) | | | | (126.1) |
| Observations | 566,035 | 566,035 | 566,035 | 566,035 | 566,035 | 566,035 | 566,035 | 566,035 |
| R-squared | 0.210 | 0.222 | 0.215 | 0.234 | 0.225 | 0.237 | 0.230 | 0.249 |
| N clusters | 37765 | 37765 | 37765 | 37765 | 37765 | 37765 | 37765 | 37765 |

| | Individual Earnings — Percentile Rank | | | | | | | |
|---|---|---|---|---|---|---|---|---|
| Agreeableness PGI Proband | -0.175 | -0.145 | -0.148 | -0.193 | -0.317 | -0.289 | -0.292 | -0.332 |
| | (0.182) | (0.180) | (0.181) | (0.178) | (0.180) | (0.178) | (0.179) | (0.177) |
| Conscientiousness PGI Proband | 0.255 | -0.0913 | 0.277 | 0.112 | 0.346 | 0.00151 | 0.373 | 0.209 |
| | (0.183) | (0.181) | (0.182) | (0.179) | (0.182) | (0.180) | (0.181) | (0.178) |
| Extraversion PGI Proband | 0.150 | 0.0940 | -0.0386 | 0.0755 | 0.307 | 0.260 | 0.123 | 0.232 |
| | (0.180) | (0.178) | (0.180) | (0.176) | (0.179) | (0.177) | (0.179) | (0.175) |
| Emotional Stability PGI Proband | 1,335 | 1,258 | 1,293 | 0,747 | 0,614 | 0,519 | 0,561 | 0,0701 |
| | (0.184) | (0.182) | (0.183) | (0.180) | (0.197) | (0.195) | (0.196) | (0.194) |
| Openness PGI Proband | 0.133 | 0.161 | 0.119 | 0.186 | -0.387 | -0.388 | -0.416 | -0.368 |
| | (0.183) | (0.181) | (0.182) | (0.178) | (0.187) | (0.186) | (0.186) | (0.183) |
| Agreeableness PGI Parents | 0.129 | 0.171 | 0.163 | 0.121 | 0.0989 | 0.146 | 0.138 | 0.0977 |
| | (0.183) | (0.182) | (0.182) | (0.180) | (0.182) | (0.180) | (0.181) | (0.178) |
| Conscientiousness PGI Parents | -0.0243 | -0.0518 | -0.0554 | -0.0520 | -0.0177 | -0.0447 | -0.0471 | -0.0455 |
| | (0.182) | (0.180) | (0.181) | (0.178) | (0.181) | (0.179) | (0.180) | (0.177) |
| Extraversion PGI Parents | -0.214 | -0.210 | -0.180 | -0.202 | -0.147 | -0.145 | -0.116 | -0.137 |
| | (0.182) | (0.179) | (0.181) | (0.178) | (0.181) | (0.178) | (0.180) | (0.177) |
| Emotional Stability PGI Parents | -0,0955 | -0,0846 | -0,12 | -0,136 | -0,385 | -0,351 | -0,377 | -0,388 |
| | (0.183) | (0.181) | (0.182) | (0.179) | (0.197) | (0.194) | (0.196) | (0.192) |
| Openness PGI Parents | 0.370 | 0.355 | 0.381 | 0.313 | 0.229 | 0.208 | 0.242 | 0.174 |
| | (0.183) | (0.182) | (0.183) | (0.179) | (0.189) | (0.187) | (0.188) | (0.185) |
| EA PGI Proband | | | | | 2.268 | 2.336 | 2.313 | 2.382 |
| | | | | | (0.218) | (0.217) | (0.218) | (0.214) |
| EA PGI Parents | | | | | 0.561 | 0.548 | 0.510 | 0.510 |
| | | | | | (0.219) | (0.217) | (0.219) | (0.216) |
| Cog. Performance PGI Proband | | | | | 0.577 | 0.663 | 0.606 | 0.534 |
| | | | | | (0.210) | (0.208) | (0.209) | (0.206) |
| Cog. Performance PGI Parents | | | | | -0.0863 | -0.0668 | -0.0577 | -0.0697 |
| | | | | | (0.212) | (0.209) | (0.211) | (0.208) |
| MDD PGI Proband | | | | | -0.772 | -0.758 | -0.774 | -0.639 |
| | | | | | (0.194) | (0.192) | (0.193) | (0.190) |
| MDD PGI Parents | | | | | -0.397 | -0.336 | -0.324 | -0.316 |
| | | | | | (0.194) | (0.192) | (0.194) | (0.191) |
| Conscientiousness Phenotype | | 3.049 | | | | 3.136 | | |
| | | (0.117) | | | | (0.116) | | |
| Extraversion Phenotype | | | 2.121 | | | | 2.129 | |
| | | | (0.118) | | | | (0.117) | |
| Emotional Stability Phenotype | | | | 4,292 | | | | 4,267 |
| | | | | (0.117) | | | | (0.116) |
| Observations | 566,035 | 566,035 | 566,035 | 566,035 | 566,035 | 566,035 | 566,035 | 566,035 |
| R-squared | 0.190 | 0.203 | 0.197 | 0.215 | 0.205 | 0.219 | 0.211 | 0.230 |
| N clusters | 37765 | 37765 | 37765 | 37765 | 37765 | 37765 | 37765 | 37765 |

| | Net Wealth — Levels (€) | | | | | | | |
|---|---|---|---|---|---|---|---|---|
| Agreeableness PGI Proband | -2,755 | -2,625 | -2,686 | -2,803 | -4,001 | -3,879 | -3,944 | -4,039 |
| | (2,231) | (2,229) | (2,231) | (2,229) | (2,224) | (2,222) | (2,223) | (2,221) |
| Conscientiousness PGI Proband | 4,279 | 2,787 | 4,333 | 3,860 | 3,387 | 1,919 | 3,447 | 3,000 |
| | (2,211) | (2,215) | (2,210) | (2,209) | (2,204) | (2,207) | (2,203) | (2,202) |
| Extraversion PGI Proband | 82.86 | -162.0 | -396.4 | -140.1 | 186.0 | -20.41 | -229.9 | -29.52 |
| | (2,217) | (2,214) | (2,222) | (2,214) | (2,200) | (2,197) | (2,205) | (2,197) |
| Emotional Stability PGI Proband | 6.158 | 5.828 | 6.050 | 4.432 | 578,6 | 182,1 | 458,7 | -954,3 |
| | (2,245) | (2,241) | (2,244) | (2,255) | (2,408) | (2,406) | (2,407) | (2,416) |
| Openness PGI Proband | -1,242 | -1,118 | -1,276 | -1,082 | -2,253 | -2,251 | -2,317 | -2,197 |
| | (2,165) | (2,161) | (2,165) | (2,162) | (2,232) | (2,228) | (2,232) | (2,229) |
| Agreeableness PGI Parents | 884.0 | 1,065 | 969.2 | 860.3 | -130.4 | 70.29 | -41.28 | -134.4 |
| | (2,249) | (2,247) | (2,249) | (2,246) | (2,243) | (2,242) | (2,243) | (2,241) |
| Conscientiousness PGI Parents | 2,002 | 1,887 | 1,924 | 1,917 | 2,289 | 2,178 | 2,223 | 2,207 |
| | (2,224) | (2,221) | (2,222) | (2,220) | (2,210) | (2,207) | (2,209) | (2,207) |
| Extraversion PGI Parents | -1,238 | -1,220 | -1,151 | -1,199 | 283.7 | 297.2 | 354.0 | 314.9 |
| | (2,199) | (2,197) | (2,199) | (2,197) | (2,184) | (2,182) | (2,184) | (2,182) |
| Emotional Stability PGI Parents | 8.041 | 8.085 | 7.980 | 7.916 | 1.522 | 1.659 | 1.541 | 1.508 |
| | (2,193) | (2,189) | (2,192) | (2,191) | (2,360) | (2,356) | (2,359) | (2,358) |
| Openness PGI Parents | 1,012 | 945.4 | 1,041 | 843.9 | -3,024 | -3,121 | -2,994 | -3,180 |
| | (2,178) | (2,175) | (2,178) | (2,176) | (2,241) | (2,239) | (2,241) | (2,239) |
| EA PGI Proband | | | | | 14,239 | 14,532 | 14,342 | 14,571 |
| | | | | | (2,617) | (2,613) | (2,617) | (2,613) |
| EA PGI Parents | | | | | -7,247 | -6,892 | -7,182 | -7,374 |
| | | | | | (2,485) | (2,482) | (2,485) | (2,481) |
| Cog. Performance PGI Proband | | | | | -10,733 | -10,670 | -10,737 | -10,356 |
| | | | | | (2,416) | (2,412) | (2,416) | (2,414) |
| Cog. Performance PGI Parents | | | | | 20,364 | 20,307 | 20,247 | 20,217 |
| | | | | | (2,691) | (2,686) | (2,692) | (2,687) |
| MDD PGI Proband | | | | | -4,192 | -4,103 | -4,129 | -4,140 |
| | | | | | (2,515) | (2,511) | (2,514) | (2,511) |
| MDD PGI Parents | | | | | -8,535 | -8,276 | -8,372 | -8,306 |
| | | | | | (2,363) | (2,361) | (2,361) | (2,362) |
| Conscientiousness Phenotype | | 13,100 | | | | 13,313 | | |
| | | (1,360) | | | | (1,349) | | |
| Extraversion Phenotype | | | 5,353 | | | | 4,775 | |
| | | | (1,406) | | | | (1,396) | |
| Emotional Stability Phenotype | | | | 12.645 | | | | 12.064 |
| | | | | (1,349) | | | | (1,336) |
| Observations | 558,346 | 558,346 | 558,346 | 558,346 | 558,346 | 558,346 | 558,346 | 558,346 |
| R-squared | 0.091 | 0.093 | 0.091 | 0.092 | 0.104 | 0.106 | 0.104 | 0.106 |
| N clusters | 37766 | 37766 | 37766 | 37766 | 37766 | 37766 | 37766 | 37766 |

| | Net Wealth — Percentile Rank | | | | | | | |
|---|---|---|---|---|---|---|---|---|
| Agreeableness PGI Proband | -0.0761 | -0.0596 | -0.0732 | -0.0813 | -0.224 | -0.209 | -0.222 | -0.228 |
| | (0.201) | (0.200) | (0.200) | (0.200) | (0.199) | (0.198) | (0.199) | (0.198) |
| Conscientiousness PGI Proband | 0.510 | 0.320 | 0.513 | 0.466 | 0.416 | 0.229 | 0.418 | 0.375 |
| | (0.197) | (0.198) | (0.197) | (0.197) | (0.196) | (0.196) | (0.196) | (0.196) |
| Extraversion PGI Proband | -0.160 | -0.191 | -0.181 | -0.184 | -0.134 | -0.160 | -0.149 | -0.157 |
| | (0.198) | (0.198) | (0.199) | (0.198) | (0.196) | (0.196) | (0.196) | (0.196) |
| Emotional Stability PGI Proband | 0,707 | 0,665 | 0,702 | 0,523 | 0,00338 | -0,0469 | -0,000797 | -0,159 |
| | (0.200) | (0.199) | (0.200) | (0.201) | (0.215) | (0.214) | (0.215) | (0.215) |
| Openness PGI Proband | -0.367 | -0.351 | -0.368 | -0.350 | -0.492 | -0.491 | -0.494 | -0.486 |
| | (0.196) | (0.196) | (0.196) | (0.196) | (0.202) | (0.201) | (0.202) | (0.201) |
| Agreeableness PGI Parents | -0.159 | -0.136 | -0.156 | -0.162 | -0.260 | -0.235 | -0.257 | -0.261 |
| | (0.200) | (0.199) | (0.200) | (0.199) | (0.198) | (0.198) | (0.198) | (0.198) |
| Conscientiousness PGI Parents | 0.302 | 0.287 | 0.298 | 0.293 | 0.339 | 0.325 | 0.336 | 0.330 |
| | (0.198) | (0.198) | (0.198) | (0.198) | (0.196) | (0.196) | (0.196) | (0.196) |
| Extraversion PGI Parents | -0.198 | -0.196 | -0.195 | -0.194 | -0.0442 | -0.0425 | -0.0418 | -0.0409 |
| | (0.199) | (0.199) | (0.199) | (0.199) | (0.197) | (0.196) | (0.197) | (0.197) |
| Emotional Stability PGI Parents | 0,733 | 0,738 | 0,73 | 0,719 | 0,0544 | 0,0718 | 0,055 | 0,0529 |
| | (0.199) | (0.199) | (0.199) | (0.199) | (0.214) | (0.214) | (0.214) | (0.214) |
| Openness PGI Parents | 0.00701 | -0.00144 | 0.00827 | -0.0109 | -0.355 | -0.368 | -0.354 | -0.372 |
| | (0.196) | (0.196) | (0.196) | (0.196) | (0.202) | (0.202) | (0.202) | (0.202) |
| EA PGI Proband | | | | | 1.658 | 1.695 | 1.661 | 1.693 |
| | | | | | (0.237) | (0.236) | (0.237) | (0.237) |
| EA PGI Parents | | | | | 1.811 | 1.803 | 1.806 | 1.795 |
| | | | | | (0.240) | (0.239) | (0.240) | (0.239) |
| Cog. Performance PGI Proband | | | | | -0.758 | -0.713 | -0.755 | -0.771 |
| | | | | | (0.224) | (0.224) | (0.224) | (0.224) |
| Cog. Performance PGI Parents | | | | | -0.269 | -0.258 | -0.267 | -0.264 |
| | | | | | (0.226) | (0.225) | (0.226) | (0.226) |
| MDD PGI Proband | | | | | -1.363 | -1.355 | -1.363 | -1.323 |
| | | | | | (0.212) | (0.212) | (0.212) | (0.212) |
| MDD PGI Parents | | | | | -0.955 | -0.922 | -0.949 | -0.931 |
| | | | | | (0.212) | (0.211) | (0.212) | (0.212) |
| Conscientiousness Phenotype | | 1.667 | | | | 1.689 | | |
| | | (0.124) | | | | (0.122) | | |
| Extraversion Phenotype | | | 0.229 | | | | 0.167 | |
| | | | (0.127) | | | | (0.125) | |
| Emotional Stability Phenotype | | | | 1,347 | | | | 1,275 |
| | | | | (0.126) | | | | (0.124) |
| Observations | 558,346 | 558,346 | 558,346 | 558,346 | 558,346 | 558,346 | 558,346 | 558,346 |
| R-squared | 0.006 | 0.010 | 0.006 | 0.008 | 0.025 | 0.029 | 0.025 | 0.027 |
| N clusters | 37766 | 37766 | 37766 | 37766 | 37766 | 37766 | 37766 | 37766 |